# One-Pot Printing of Robust Multimaterial Gradients

*Sijia Huang\*, Steven M. Adelmund, Pradip S. Pichumani, Johanna J. Schwartz, Yiğit Mengüç, Maxim Shusteff, Thomas J. Wallin\**

Dr. J. J. Schwartz, Dr. M. Shusteff
Lawrence Livermore National Laboratory,
Livermore, CA 94550, USA

Dr. S. Huang,
Department of Chemical and Biochemical Engineering
Univeristy of Colorado-Boulder, Boulder, CO 80303, USA
Current address:
Lawrence Livermore National Laboratory, Livermore, CA 94550, USA
Email: huang55@llnl.gov

Dr. Y. Mengüç
Collaborative Robotics and Intelligent Systems (CoRIS) Institute,
Oregon State University, Corvallis, OR 97331, USA

Dr. S. Huang, Dr. S. Adelmund, P. Pichumani, Dr. Y. Mengüç, Dr. T. J. Wallin
Reality Labs at Meta, Redmond, WA 98052, USA
Email: thomaswallin@fb.com

Seamless multimaterial construction, particularly joining soft, stretchable tissues with stiff, inextensible structures, is a common motif in animal physiology. Such continuous mechanical gradients remain challenging to reproduce in engineered systems as current resin chemistries typically result in a single fixed set of properties. As an alternative to single-property materials, we introduce a ternary sequential reaction scheme that produces multimaterials by profoundly altering the polymer microstructure from within a single resin composition. In this system, the photodosage during 3D printing sets both the shape and extent of conversion for each subsequent reaction. The different polymerization mechanisms of the subsequent stages allow our single photochemistry to exhibit a diverse range of soft (Young's Modulus, $E \sim 400$ kPa; ultimate elongation, $dL/L_0 \sim 300\%$) and stiff ($E \sim 1.6$ GPa; $dL/L_0 \sim 3\%$), providing the capability to match the mechanical properties of commercial polymers and biological tissues. Further, we successfully pattern photostable and mechanically robust modulus gradients ($d[E_{r,stiff}/E_{r,soft}]/dx > 1000$ mm$^{-1}$) that exceed those found in squid beaks and human knee entheses. We demonstrate the ability to 3D print intricate multimaterial architectures by fabricating a soft, wearable braille display.

1. Introduction

Across length scales, biology combines soft and stiff structures to control mechanical performance and enable higher functionality. As a prime example, load-bearing activity in vertebrates relies on the connection of rigid bones and soft muscle tissue[1]. Similarly, biomedical devices, soft robotics, human-computer interaction (HCI), and other fields seek to revolutionize technology by mechanically coupling the soft, extensible tissues of the body to highly functional, but rigid components (motors, sensors, microelectronics, batteries, etc) of conventional engineered systems. These efforts require new materials and manufacturing developments that enable robust mechanical connections between such disparate materials[2].

Despite building with materials that span over nine orders of magnitude in modulus ($E$)[3], biology rarely utilizes abrupt mechanical transitions as continuous gradients are more failure resistant (e.g., optimum strength and maximum flaw tolerance under complex load). Unfortunately, synthetic systems often fail to capture such smooth gradients between soft and stiff components at similar length scales[4–6]. Emerging additive manufacturing, or 3D printing, technologies can potentially close the gap in structural complexity between human-made and natural systems, but these fabrication methods still struggle to produce robust synthetic multimaterials or mechanical gradients comparable to those found in physiology ($d[E_{stiff}/E_{soft}]/dx > 100$ mm$^{-1}$)[7].

In general, 3D printing strategies either selectively dispense a material ink on to a build stage prior to solidification or selectively irradiate a vat of precursors (resin or powder) to induce solidification[7,8]. Such conventional 3D printing often produces "single property materials," i.e. each printed material corresponds to a specific precursor composition. Dispensing methods (e.g. direct ink writing [DIW], inkjetting, etc) can switch between multiple ink heads or dynamically mix precursors from different reservoirs to realize multimaterial capabilities[9]. Yet, due to the practicalities of fluid flow, these strategies encounter significant obstacles to achieving the structure complexity and compositional precision necessary to mimic the complex design found in biology[5,9–11]. Further, material pairings that possess disparate moduli (i.e., $E_{stiff}/E_{soft} > 100$) are often very chemically dissimilar. This dissimilarity can introduce two challenges: i) co-processability is particularly problematic when regulating the flow of inks with divergent rheological behavior and ii) obtaining strong interfacial bonds between materials requires further reactive compatibility. Alternatively, vat-based strategies (e.g. selective laser sintering [SLS], digital light processing [DLP][12], volumetric additive manufacturing [VAM][13], etc.) transmit electromagnetic radiation (IR, UV, or visible light) to rapidly ($\leq 10$ m$^3$ h$^{-1}$) print intricate structures at micron resolution[13]. However, it is difficult to dynamically change the precursor chemistry for multimaterial printing. The most common strategy involves replacing the entire vat for different materials in a given layer; obtaining smooth mechanical transitions through such techniques would require a vast number of material reservoirs and become prohibitively difficult to scale[14–17].

Rather than altering the local composition of precursors, other efforts seek to produce varying stiffness within a single photochemistry by changing the processing conditions during printing. Within a given polymer backbone above the glass transition temperature ($T_g$), the modulus is directly proportional to the volumetric crosslink density ($v$). Thus, it is easy to control the extent of reaction conversion, and consequently, the apparent modulus, in acrylate photopolymers by simply limiting the photoirradiation dosage[18,19]. However, when compared to natural tissues, the demonstrated stiffness gradients available from these materials are relatively shallow ($E_{stiff}/E_{soft}$ ~ 10)[19]. Larger moduli differences ($10 < E_{stiff}/E_{soft} < 10^4$) are possible by formulating resins that contain multiple curing chemistries. Here, different crosslinking reactions (e.g. epoxy and acrylate homopolyerization[20–22]; thiol-Michael addition and acrylate hompolymerization[23,24], acrylate homopolymerization with aza-Michael and epoxy-amine addition[25]) occur in response to distinct stimuli (i.e., dual wavelengths or combined light and heat) to allow for localization of desired properties. However, regardless of strategy, these chemistries all result in materials that are not suitable for consistent, long-term consumer use. Specifically, the "soft" regions within such systems are intentionally "underpolymerized" to minimize crosslinking density. After printing, these voxels contain unreacted acrylates or epoxides that continue to crosslink and stiffen under even modest exposure to ambient light or heat. Post-processing can remove some unbound, e.g. completely unreacted, species from the polymer network, but the corresponding loss in mass will result in undesired changes to both the shape and stiffness of the object[26]. Beyond the poor long term stability, these partially cured networks are often not stretchable (ultimate elongation, d$L$/$L_0$ <40%) or tough, which precludes their use in elastomeric devices.

Here, we introduce a sequential thiol-ene-epoxy framework as a new framework for creating stable multimaterials capable of mimicking a wide range of mechanical performance available in natural tissues and commercial polymers. Specifically, our approach utilizes photoinitiated thiol-ene reactions followed by thiol-epoxy step growth polymerization and then epoxy homopolymerization at elevated temperatures. This unique design creates a reactive sequence where the photoirradiation dosage during printing determines the thiol and ene functional group conversion ($x$) in the first stage, which both sets the object shape and determines the amount of "soft" thiol-epoxy network created in the second stage, which consequently determines the amount of "stiff" epoxy network formed in the third stage. Thus, the applied photodosage directly dictates the extent of "soft" and "stiff" segments within the printed part. This precise control yields a diverse range of polymers from soft elastomers ($E_{soft}$ ~400 kPa, d$L/L_0$ ~300%) to glassy thermosets ($E_{stiff}$ ~1.6 GPa, d$L/L_0$ ~ 3%). Unlike other photopatterned multimaterials, these sequential reactions always result in the full consumption of readily reacted groups during processing which imparts environmental stability ($E_{stiff}/E_{soft} > 10^3$ after the equivalent of ~1,000 h of solar irradiation) to our final material. Not only do these mechanical properties approximate those found in commodity polymers (acrylics, polyurethanes, silicones, hydrogels) and anatomy (tendons, ligaments, skin, gastrointestinal tissues) but the photodosage controlled, continuous stiffness gradients ($0 \leq $ d$[E_{r,stiff}/E_{r,soft}]/$d$x \leq $ 1300 mm$^{-1}$) produced from our system can approximate those found in healthy human knee entheses and squid beaks. To demonstrate the capabilities that such biomimetic photopatterned multimaterials offer engineered systems, we use this chemistry to rapidly ($t_{print}<$ 10 min) 3D print a soft wearable braille display.

## 1. Results

**Chemical Design of Ternary Sequential Photopolymer Resin**

Designing a materials chemistry that can selectively produce soft and stiff materials requires consideration of the phenomenological origin of the mechanical properties. Solid polymeric materials are ensembles of macromolecular chains held together by both intra- and interchain interactions. At the microscale, the density and relative strength of these interactions determine how the polymer network deforms under a given load which, in turn, dictates the observed macro-scale mechanical properties. Thus, polymer properties highly depend on both backbone structure and chain arrangements. As a result, creating a stable, one-pot multimaterial chemistry is particularly challenging: the initial chemical composition of the formulation cannot change and intentionally leaving unreacted precursors threatens long-term stability (e.g. the reactions may proceed after processing to alter properties). Instead, we leverage different polymerization mechanisms to yield drastically different crosslinking densities, viscoelastic properties, and therefore, mechanical performance.

Our resin (**Figure 1a**) combines a triallyl (-ene) species, a thiol mixture (9:1 molar ratio of di- and tetra-thiol molecules), and a diepoxy. For simplicity, we use a 1:1:1 stoichiometric ratio between ene, thiol, and epoxide functional groups. This composition creates a three-stage curing chemistry: two step growth reactions (thiol-ene and thiol-epoxy polymerizations) and a chain growth polymerization (epoxy homopolymerization). As shown in **Figure 1b**, the first stage is a radical initiated thiol-ene photopolymerization that rapidly forms a loosely crosslinked, percolated network during printing. The second stage continues to build a soft network based on the step growth thiol-epoxy polymerizations initiated by a thermally latent imidazole (Technicure® LC-80) under modest temperatures (T ~ 80°C). During the higher temperature (T ~ 120°C) third stage, the anionic homopolymerization of the rigid diepoxy species creates a stiff, highly crosslinked network, as reported in previous literature[27]. Key to this design is that this sequential reaction is effectively controllable—we can create separate curing "stages" during processing by simply applying different stimuli or leveraging differential reaction rates.

The proposed stoichiometry and sequence of these reactions enables precise control over each stage of curing. As shown in **Figure 1c**, the photoexposure dosage ($H_e$) applied during printing dictates the % conversion ($x$) of ene and thiol functional groups. In the second stage, the remaining mol fraction (1-$x$) of thiols act as a limiting reagent and reacts stoichiometrically with epoxide functional groups (i.e., 1:1 thiol to epoxide) through stepwise addition. Lastly, during the third stage, the residual epoxides ([1-[1-$x$] = $x$) homopolymerize in a chain growth fashion. Thus, our chemical design suggests that photoexposure can not only impart the geometric shape of the printed object but ultimately dictate the local network microstructure and, consequently, material properties.

**Reaction Kinetics during Processing**

We use Fourier-transform infrared (FT-IR) spectroscopy to infer the consumption of chemical groups over time (ene, thiol, and epoxide) and verify the control over the proposed three reactions (see *Materials and Methods* and **Figure S1-S2** for more information). As shown in Figure 1c, the conversion of thiol and ene groups, $x$, depends on the applied photodosage ($H_e$) as described by the typical first-order reaction kinetic model shown in Equation (1)[28].

$$[x = 1 - e^{-0.02H_e}] \qquad \text{Equation (1)}$$

Note that termination steps (radical recombination and chain transfer) rapidly quench any free radicals as evidenced by the thiol-ene conversion quickly plateauing after cessation of light (see **Figure S3**)[29]. Thus, control of the light engine (e.g. LED projector) allows for spatiotemporal control of the chemical reaction.

In **Figure 1d-f**, the photodosage is varied during the first reactive stage followed by thermal curing at 80 °C and 120 °C. In the absence of light irradiation (Figure 1d), no thiol-ene addition occurs ($x \sim 0$) which enables nearly all thiols to undergo nucleophilic-addition to epoxides, in the second step. Consequently, only a few remaining epoxides are available to undergo homopolymerization in the third stage. As shown in Figure 1e, at intermediate photodosages ($H_e$ = 62.5 mJ cm$^{-2}$), the photopolymerization consumes thiol and ene groups evenly ($x \sim 0.52$). As expected, the residual thiols serve as a limiting reagent and consume equimolar epoxide functional groups in the second stage. The epoxide conversion appears to increase slightly even after the thiol conversion approaches 1. This increase suggests additional reactions (i.e. epoxy-epoxy homopolymerization) can occur, albeit more slowly, at this temperature. However, differential scanning calorimetry (DSC) (see **Figure S4**) confirms that the thiol-epoxide reaction remains dominant until full consumption of thiols— thiol-epoxy reactions are almost 10X faster than epoxide homopolymerization at 80°C, which is consistent with previous literature[30]. At the elevated temperatures of the third stage, the epoxy homopolymerization accelerates and approaches full conversion. Alternatively, a longer first-stage photoexposure ($H_e$ = 175 mJ cm$^{-2}$) consumes 98% of the ene and thiol groups (Figure 1f). With few remaining thiols, the second stage thiol-epoxy addition is minimal, and nearly all epoxides homopolymerize during the final step. In total, these experiments, as well as photo- and thermal rheology (see **Figure S5**), indicate the sequential framework behaves as expected.

**Mechanical Properties, Gel Fraction, and Aging in Multimaterial Resins**

After demonstrating the ternary reaction motif, we characterized the tensile properties resulting from different initial photodosages (see *Materials and Methods*). **Figure 2a** contains the average (N ≥ 7) stress-strain plots for each photodosages (note the log-linear scaling). No photodsage ($H_e$ ~ 0 mJ·cm$^{-2}$), when only the loosely crosslinked thiol-epoxy network forms ($x$ ~ 0), produces a soft ($E$ ~ *410 kPa*), stretchable material with ~300% elongation (d$L$/$L_0$). In contrast, at $H_e \geq 250$ mJ·cm$^{-2}$ the resin sequentially forms a thiol-ene network ($x$ ~ 1) interpenetrated by a rigid epoxy homopolymer. When compared to the $x$ ~ 0 variant, this material is over 3800 times stiffer ($E$ ~ 1.6 GPa) with 100 times decrease in ultimate elongation, (d$L$/$L_0$ ~3%). The resin exhibits intermediate behavior at the other photodosages—in general, as the applied photodosage and conversion ($x$) increases, the final material becomes stiffer and less stretchable.

Our single resin chemistry exhibits a profound diversity in mechanical properties that span a range comparable to commercial polymers and soft biological tissues. **Figure 2b** contextualizes this ternary sequential chemistry by comparing the modulus and ultimate elongation of biomaterials, commonly used polymers, and previous multimaterial photochemistries. Our stiffest material ($H_e$= 220 mJ cm$^{-2}$) shows a Young's modulus, $E$ ~ 1.6 GPa and fracture strain similar to polymethyl methacrylate (PMMA[31] or acrylic[32]) and polystyrene (PS)[33], whereas the intermediate and softer materials transit a regime of mechanical performance that includes polyurethanes (Smooth-Cast 45D)[34], silicones (Sylgard 184)[35], and hydrogels (gelatin methacryloyl hydrogel)[36]. Similarly, the available properties in our thiol-ene-epoxy chemistry cover a comparable range to that of tendons[37], ligaments[38], skin[39,40], and gastrointestinal tissues[41,42]. While the acrylate-amine-epoxy photochemistry also exhibits a short-lived (see below) three order of magnitude change in stiffness immediately after processing, the ultimate elongation for this multimaterial remains consistently low (d$L$/$L_0$~30%), which hinders biomedical and soft robotic applications that require highly extensible components. From this perspective, our photochemistry provides for the greatest diversity in mechanical properties and greatly enlarges the design space for printed devices.

Interestingly, conventional theories do not fully explain the range of observed modulus ratio ($E/E_{\text{soft}}$) in our system. We can attempt to model the multimaterial as a composite of the individual polymer networks (thiol-ene, thiol-epoxy, and epoxy-epoxy[43]) whose properties are included in Figure 2b. However, **Figure 2c** shows that neither an isostress ($E/E_{\text{soft, max}}$ ~ 2) or isostrain ($E/E_{\text{soft, max}}$ ~ 40) composite model predicts the modulus ratio as a function of 1$^{\text{st}}$ stage conversion, $x$ (see SI section 1 for these calculations). Alternatively, the affine network model holds that modulus (Young's modulus, $E$, or storage modulus, $E'$) is proportional to crosslink density, $v^{44}$. Fortunately, we can change the crosslink density within the same composition because the diepoxy functionality, or bonds formed per BisDE molecule, changes depending on whether the monomer participates in thiol-epoxy addition (f=2) or epoxy homopolymerization (f=4). Thus, we estimate $v$ as a function of $x$, (see SI Section 2 for more detail), but this only predicts a $E/E_{soft, max}$ ~ 15. Since the relationship between modulus and crosslink density only holds for materials in their rubbery regime, we attribute this excess modulus ratio between our soft and stiff material to the temperature dependent changes in the molecular mobility of the polymer chains. As shown in **Figure 2d** and **Figure S6**, the glass transition temperature ($T_g$) of our materials depends on the applied photodosage. For the loosely crosslinked thiol-epoxy material ($H_e = 0$ mJ·cm$^{-2}$), the glass transition is $T_g$ ~ -9 °C. By comparison, high photodosage ($H_e \geq 220$ mJ·cm$^{-2}$) favors the formation of a densely crosslinked epoxy network with a corresponding glass transition above room temperature ($T_g$ ~35 °C). Supporting our hypothesis, when we measure the storage moduli at 100°C ($T>>T_g$), all of the materials are rubbery and the observed moduli ratio moves closer to the expectations of the affine network model (blue squares in **Figure 2c**). Beyond mechanical performance, the applied photodosage also impacts other material properties. As shown in **Figure S1**, the thiol-epoxy addition leaves a residual hydroxyl (-OH) side group while epoxy-epoxy reaction imparts an ether (R-O-R) group onto the chain. This chemical difference manifests itself as a 25% decrease in surface energy from $x$ ~ 0 to $x$~1 (see **Figure S7**) which has profound implications for adhesion.

A key distinction between the thiol-ene-epoxy resin and other multimaterial photochemistries is that our fully processed multimaterial does not contain reactive species in the soft regions, which ensures stable mechanical properties under ambient light and temperature[19,23–25]. To test this hypothesis, fully processed samples underwent additional photoexposure ($H_e = 10$ J·cm$^{-2}$, $\lambda = 405$ nm) and heat treatment (120 °C, 1 h). As shown in **Figure 2e**, the softest acrylate-amine-epoxy material undergoes significant aging ($E_{soft,initial}$ ~ 0.8 MPa to $E_{soft, aged}$ ~ 470 MPa). While the solar irradiance varies significantly depending on the time of day, date, geographic position, altitude, local weather conditions, etc., we note that sea-level irradiance in Northern latitudes still frequently exceed 1 mW cm$^{-2}$ at $\lambda = 405$ nm (see **Figure S8a**) Thus, the acrylate-amine-epoxy material suffers a ~125x loss of gradients in under 10 J·cm$^{-2}$, which can be equivalent to less than 5 mins of ambient sunlight. Not only does such a rapid, drastic change in the stress-strain response of a material threaten reliable device operation, but aging while under strain may also contribute to permanent shape change, as seen in **Figure S8b**. By comparison, our thiol-ene-epoxy system maintains a 1000x moduli gradient at dosages up to 5,000 J·cm$^{-2}$—approximately 1150 hours of solar irradiance at $\lambda = 405$ nm—indicating the robust stability of this system (see **Figure S8c** for the data from materials of intermediate moduli). These results suggest that object made from our chemistry would exhibit longer performance lifecycles in real-world applications.

We acknowledge that our softest material ($x = 0$) ages slightly under photoexposure and should possess unreacted ene groups from the first stage. However, we attribute the change in properties to the slow evaporation of unbound ene monomers and not their homopolymerization (see Supplemental Information for more details). **Figure 2f** shows the gel fraction (GF), or proportion of material (by mass) that cannot be separated from as-printed polymer network. For our thiol-ene-epoxy system, the GF measurements agree with estimates for removal of fully unreacted ene monomers as shown in Equation (2) (see SI for derivation).

$$GF_{expected} = 1 - (1 - x)^f \, m_{ene} \qquad \text{Equation (2)}$$

Where $x$ is the ene functional group conversion, $f$ is the functionality of the ene monomer ($f = 3$), and $m_{ene}$ is the mass fraction of the ene monomer in the resin ($m_{ene} = 0.22$). Since the GF remains constant after UV aging ($H_{e,\,aged} = 10$ J·cm$^{-2}$, $\lambda = 405$ nm), these experiments further suggest that ene homopolymerization in our system is improbable.

To mitigate aging, many of previous dual-stage multimaterial photochemisties use dialysis to extract residual reactive groups from the "gelled material" and minimize post-print polymerization[21,45]. As evident in Figure 2F, the soft materials from these previous systems (i.e, those polymerized by $\lambda_{soft}$) possess a small GF (<40%) which creates obstacles for high resolution, multimaterial 3D printing. The removal of such large quantities of unbound species results in a change of volume unless the solvation process is appropriately chosen to enable extraction without causing collapse or swelling of the gel—essentially unfeasible as the local composition, crosslinking, and stiffness of the polymer network varies[20,46]. Modifications to the 3D printed design can attempt to accommodate such changes in volume, but this becomes mathematically challenging for multimaterial objects as gel fraction, and therefore shrinkage, varies by voxel. Additionally, extraction will be less effective at minimizing UV aging for resins based on high functionality (f >> 2) or heterofunctional monomers (e.g. glycidyl acrylate) that tether into the system even when only partially reacted during printing. This limitation is evident in the softest acrylate-amine-epoxy material aging significantly ($E_{aged}/E_0 > 500$) despite its high gel fraction (~80%)[25].

**Shape Resolution and Mechanical Property Resolution**

To realize the full potential of multimaterial additive manufacturing, it must be possible to pattern both shape and mechanical performance of our thiol-ene-epoxy materials at high resolution. Unfortunately, a given resin chemistry will exhibit drastically different shape resolution depending on the material hardware and process parameters. Further, for vat polymerization, the attenuation (absorption and scattering) of light through the resin produces different resolutions for features parallel ($z$) and normal ($x$-$y$) to the direction of incident light[47,48]. Still, to benchmark the $x$-$y$ resolution, we subject our resin to a single 3 mm wide photoexposure ($H_e$ ~ 300 mJ·cm$^{-2}$, $\lambda$ = 405 nm) prior to washing with IPA to remove the ungelled material. Once printed, we measure the heights of the corresponding cylinders using 3D profilometry (Keyence VK-3200 G2 series microscope). **Figure 3a** depicts the average height profile of scans along the $x$ axis at different $y$ values (N = 8). Note that at the profile gradually narrows and eventually rounds for large $z$ values. These observations are consistent with light attenuation (scattering and absorption) and the inhibition of polymerization by radical scavengers in the resin (dissolved $O_2$ and pyrogallol)[47,48]. Most vat polymerization processes involving commercial resins target layer heights of approximately 100 μm. At $z$ = 100 μm above the resin-window interface, the cured material possesses a width of ~3.12 μm which is within 4% of the target design and a deviation that represents only a few multiples of our light source's nominal pixel size (50 μm x50 μm).

As light passes through a medium it attenuates according to the optical properties (absorption, scattering, etc.) of that material. During photopolymerization, this phenomena results in a "cure depth" ($C_d$) or distance from the light-resin interface at which the applied photodosage falls below the critical gel dose ($H_{e,gel}$) and the resin remains uncured. Hence, simply modulating the light dosage to the proper cure $C_d$ allows for tunable z-resolution during printing Derived from Beer's Law, Equation 3 describes the mathematical relationship between cure depth and applied photodosage[49,50].

$$C_d = D_p \ln\left(\frac{H_e}{H_{e,gel}}\right) = D_p \ln(H_e) - D_p \ln(H_{e,gel}) \quad \text{Equation (3)}$$

Where $D_p$ is the penetration depth, a constant which captures the attenuation behavior of the resin. We solve this relation experimentally by exposing the resin to varying photoirradiation dosages and measuring the corresponding height of the cured object. To collect all data in a single experiment, we create a photopattern (Figure S9b) containing an array of greyscaled circles. We then infer ($C_d$) as the max height of these cylinders and plot these values as a function of the natural log of photoexposure. As shown in Figure **3b**, the resulting data is linear (R ~ 0.96). We extrapolate a critical gel dosage of $H_{e,gel}$ = 67.5 mJ cm$^{-2}$ and $D_p$ of 1.22 mm. We note that the addition of dyes to the resin can readily tune these parameters (see Figure S9). Still, $x$, $E$, and $C_d$ are all $H_e$ dependent which creates profound challenges to multimaterial 3D printing of arbitrary structures with arbitrary properties (see the **SI Section: Print Parameters and Design Considerations for Multimaterial 3D Printing** ).

To quantify the mechanical property resolution, we subject the resin once more to the photopattern in **Figure 3a**. However, in this case the sample underwent thermal curing after photoexposure without any solvent rinse. We measured local stiffness by determining the pressure displacement curve at different locations via nanoindentation (**Figure S10**). The Oliver-Pharr method extracts a reduced elastic modulus ($E_r$) from the unloading curve (see *Materials and Methods* below for more detail). Note this modulus is fundamentally different than the Young's modulus ($E$) measured via tensile tests, however, a 1000x difference still exists between the stiffest and softest materials when sampling at discrete locations. A smooth transition between stiff and soft regions occurs over ~0.6 mm (Figure 3c); this corresponds to an impressive modulus ratio gradient (d[$E_{r,stiff}/E_{r,soft}$]/d$x$) of ~ 1300 mm$^{-1}$.

Most interestingly, this mechanical transition is far less abrupt than the shape resolution of our photochemistry (Figure 3a) implies. We attribute this "mechanical blurring" to diffusion of the unbound epoxy monomer. As the thiol-ene network solidifies in the first stage, a concentration gradient develops between the increasingly epoxide-rich, photo-cured regions into the more dilute, less-polymerized areas. During the 2$^{nd}$ stage, the temperature (80°C) is well above the $T_g$ of thiol-ene network, therefore there is little to inhibit the diffusion of unbound epoxy monomers. Such diffusion occurs until the epoxide is consumed. This explanation agrees with both the sigmoidal shape of the mechanical transition and previous observations of monomer diffusion between uncured voxels in multijet 3D printing[51]. Shallower gradients are readily accessible by utilizing a grayscaled photopattern to vary pixel dosage (**Figure 3d**).

In **Figure 3e**, we compare the synthetic mechanical gradients accessible through our multimaterial photochemistry to exemplars across length scales from biology. Squids connect their soft bodies to increasingly rigid tissues in their beaks, an arrangement that allows these cephalopods to penetrate the exoskeletons and shells of its prey[52]. The coacervation of chitin within these structures produce a 20-fold change in modulus on the centimeter scale[53]. In healthy human knee entheses, i.e. the attachment point between subchondral bone and ligaments, the biomineralization of calcium alters the reduced modulus from ~70 to 700 MPa within a millimeter[54]. Our thiol-ene-epoxy material can replicate modulus ratio gradients at scales similar to these biological examples. However, atomic force microscopy (AFM) measurements suggest that the hair-like setae in geckos[55] and ladybugs[56] are orders of magnitude moduli gradients on the length scale of 10s of microns, which is beyond the capabilities of our system. These structures, which possess a gradient of the protein resilin, provide for the adhesion that allow these animals to climb smooth, vertical surfaces.

In engineered systems, a benefit of such mechanical gradients is the ability to program desired, non-uniform deformation under loads as shown in **Figure 4a**. Patterning disparate stiffnesses on opposite halves of a dogbone imparts extreme strain localizations at the interface between soft (d$L_{\text{soft}}$ ~ 340%) and stiff regions (d$L_{\text{stiff}}$ ~ 0%). As mentioned above, abrupt changes in moduli create elastic mismatches that result in stress singularities exceeding the material strength and cause premature failure[57]. However, the continuous gradients innate to our one-pot ternary chemistry mitigate this failure mode. As shown in **Figure 4b**, the multimaterial coupons (N = 8) exhibit an interfacial strength of roughly ~ 4 MPa, well above the fracture stress ($\sigma_f$) of the weaker component ($\sigma_{f,x\sim0}$ ~ 0.2 MPa). All samples failed through the body of the weaker material, away from the transition. Moreover, the tensile toughness of this multimaterial construction ($\Gamma$ ~ 4 MPa) is ~10x that of the individual materials (x=1 and x=0) (**Figure 4c**). One possible explanation for this profound enhancement in energy absorption is the sequential nature of our chemistry. The stiffest material ($x$ ~ 1) first forms a thiol-ene network that is later interpenetrated by a separate epoxy homopolymer that chemically crosslinks across the transition to the thiol-epoxy network of the soft region ($x$ ~ 0). Previous works found that such double network motifs increase fracture resistance and improve mechanical strength at multimaterial interfaces[58,59].

**3D Printed Multimaterial Braille Display**

Multimaterial construction is particularly enabling for wearable technology where engineers need to bridge the mechanical regimes of rigid, inflexible microelectronics to our highly stretchable soft tissues. The mechanically robust, environmentally stable thiol-ene-epoxy chemistry allows for the fabrication of such devices for everyday use like the multimaterial braille display shown in **Figure 5a**. Compliant features are desirable for components that interface with soft, delicate tissues of the human body, like our finger mounted band and braille pads. In other areas, stiffer components are necessary to constrain actuation (rigid cap) and define individual "dots" on the active area of the display (rigid case). A conically shaped "gradient pin" that is stiff and narrower ($E_{target}$ ~1.6 GPa, $d_{min, pin}$ ~ 1.5 mm,) at the base of the band while soft and wider ($E_{target}$ ~ 160 MPa, $d_{max, pin}$ ~ 2.25 mm) at the top mates with the smaller holes on the other side of the compliant band. We intentionally oversized the pin for light interference relative to the holes ($d_{hole}$ ~ 2 mm), creating a press fit. Numerous holes along the band permit sizing and adjustment to the individual user's anatomy.

We chose the open-source Autodesk$^{TM}$ Ember, a projection-based SLA 3D printer to fabricate our multimaterial objects (**Figure 5b**). In this bottom-up process, a single, monochromatic ($\lambda$ = 405 nm) photopattern passes through a build window at the base of a vat of resin for a set time period ($t_{exposure}$). After exposure, stepper motors translate the build stage up one-layer and the process repeats with the subsequent photopattern. Reducing the light intensity ($J_e$), or "greyscaling," specific pixels within the photopattern enables programming of the local photodosage ($H_e$). In practice, successful greyscale printing involves appropriately balancing print parameters (layer thickness, exposure time) with object design (feature dimensions, part orientation, target $E$) and resin properties (gel dosage, absorptivity). Conventional vat polymerization requires gelation of each layer during photopatterning. As a result, the available mechanical gradient is narrower than the full range accessible to 2D films. For our chemistry 1:1 stoichiometry, $f_{ene}$ ~ 3, $f_{thiol,average}$ ~ 2.2), Flory Stockmayer theory predicts a gelation at photoconversions $x > 0.65^{44}$. A linear interpolation for $E(x)$ suggests the material at this gel threshold possesses a modulus of $E$ ~ 160 MPa. However, this printable material remains ~10x less than our stiffest material which is sufficient to enable the deformation necessary for pneumatic actuation of the braille display. Refer to Figure S9 and the Supplemental Information Section 3 for a thorough treatment of these considerations and our attempts to process engineer for optimal performance.

With appropriate print parameters, we rapidly (< 10 min) print this target design before applying heat treatment for the second and third stage to set the desired mechanical properties. As shown in **Figure 5c**, the compliant band can be held into shape by the adjustment holes elastically deforming to accommodate the gradient pin. As intended, this press fit allows for ease in donning/doffing while still grounding the display to the finger to prevent slippage and loss of haptic signal[60]. After connecting pneumatic tubing (outer diameter ~ 2 mm), the braille display becomes operational. At body temperatures (~37 °C), applied pneumatic pressure causes the pads of the braille display to rise above the surface and press into the user's fingertip. As shown in **Figure 5d**, we propose integrating wrist-mounted components (WiFi receiver, replaceable $CO_2$ cartridge, pneumatic tubing, decoder and microvalve array) with our 3D printed device to enable tactile comprehension of mobile messages. Off-board valving allows for selectively addressing all six braille pads for arbitrary configurations as shown in **Supplemental Video 1** and **Figure 5e**. Applying only a modest pressure differential ($\Delta P \sim 30$ PSI) minimizes energetic requirements while still providing fast inflation and deflation ($t_{\text{on-off}} \leq 0.1$s). Though such low pressures reduces total actuation amplitude, under these conditions the inflation of individual channels is still perceptible to a user, and the rapid cycle speeds maximize the bandwidth of haptic information transfer (refresh rate ~ 10 Hz). This rate roughly corresponds to the average speed of a braille reader (~70-100 word per minute)[61].

Our 3D printed device offers many advantages when compared to the commercially available, refreshable braille displays[62]. The simplicity of a polymer-based, pneumatic device is likely far cheaper to produce (< $0.15 of resin, see **Table S1**) than alternatives based on numerous electromechanical components (>$3,500)[63]. This finger-worn haptic display also possesses a much smaller form factor, affording the user a greater degree of dexterity and portability than devices that require manually scanning fingers across a row of cells. "Static reading" of this grounded device also reduces the risk of drift errors[64]. Still, user studies would be necessary to compare content comprehension of traditional braille with a rapidly refreshing single character display. To further increase information transfer, a user could wear displays on multiple fingers, or an alternative design could increase the number of cells (e.g. 3x3 or 3x4 arrays) and actuate columns in sequence to render the equivalent sensation of gliding across a row[65]. However, both strategies add cost and complexity to the current system.

## 2. Conclusion

We present a ternary, multimaterial photochemistry based on sequential thiol-ene-epoxy reactions. The system is designed so that photoexposure dosage during the first thiol-ene stage sets both the shape during 3D printing and determines the extent of reaction for each of the subsequent thermally initiated stages. In situ tracking of chemical groups validates this controllable reactive scheme. By leveraging the mechanistic differences between the step growth (thiol-epoxy) and chain-growth (epoxy-epoxy) polymerization, we produce materials with drastically different mechanical properties that approximate commercial polymers and pattern mechanical gradients comparable to those found in biological systems. The resulting materials remain stable to UV and thermal aging after full processing.

That photoexposure dictates both the shape during printing ($H_{e,applied} > H_{e,gelation}$) and the modulus of the material [$E(H_e)$] poses issues for obtaining structures of arbitrary shape and stiffness during 3D printing (see Supplemental Information for more discussion). For example, when accounting for cure through and cumulative photoexposure, softer materials may require photoexposures near or below the critical gel dose—such parts will not survive the interlayer translations during printing. Access to softer material is still possible by entrapping partially cured resin inside thin shells of rigid material during printing[66]. As an alternative, adding ene and thiol monomers of higher functionality (f ≥ 3) would reduce the extent reaction required for gelation and permit direct access to lower conversion materials[47]. Photoinhibition, solution mask lithography, or volumetric printing strategies can further mitigate issues related to the undesired accumulation of photodosage through successive layers[67].

Despite these challenges, the development of a stable one-pot multimaterial photochemistry has profound implications for even single-material prints. The low-cost of this chemistry (< $0.30 g$^{-1}$ resin) enables on-demand manufacturing of myriad devices with varying moduli. This advantage is not only relevant for remote locations and expeditions (e.g. space exploration) where raw material transport is costly, but the simplicity of a one-pot multimaterial chemistry may also benefit traditional manufacturing environments where economy of scales make formulating and stocking a resin for each modulus between 400 kPa < E < 1.6 GPa commercially infeasible. Further, beyond this specific chemistry, this broad thiol-ene-epoxy framework offers numerous opportunities for customization—the molecular weight, functionality, and chemical composition of prepolymers all contribute to the final properties of the material. Applying the sequential reaction motif to future formulations that target other physical properties (e.g. thermal conductivity, surface energy, crystallinity, etc) would also enable opportunities for directly fabricating acoustic, mechanical, and optical metamaterials.

## Materials and Methods

*Preparation of multimaterial resins:* Pentaerythritol tetra(3-mercaptopropionate) (PETMP), triallyl-1,3,5-triazine-2,4,6-trione (TATATO), bisphenol A diglycidyl ether (BisDE), and pyrogallol (PYL) were purchased from Sigma-Aldrich. Ethylene glycol di(3-mercaptopropionate) (GDMP) was purchased from Tokyo Chemical Industry Co. A&C Catalysts, Inc kindly donated Technicure® LC-80 (LC-80). A 20/80 wt% blend of Diphenyl(2,4,6-trimethylbenzoyl)phosphine oxide/2-hydroxy-2-methylpropiophenone is used as a photoinitiator. All monomers were used as received.

A stoichiometric ratio of 1:1:1 of thiol:ene:epoxide functional groups containing 2 wt% LC-80[68], 1.25 wt% Irgacure 2022 and 0.09 wt% PYL was prepared. Table S1 provides an example recipe. To decrease the gelation threshold during the first stage and permit solidification during the second stage in the absence of 1$^{st}$ stage conversion ($x$~0), 10 mol% of thiol functional groups are from PETMP ($f$=4) while the remaining thiols belong to GDMP. A centrifugal mixer (Speedmixer DAC 600.2 VAC-LR *FlakTek*) blends resin with micro ceramic beads at 1500 rpm for 35s followed by 2000 rpm for 55s.

*Photo-Rheology:* All rheological experiments were conducted using a TA Instruments Discovery Hybrid 2 rheometer (DHR-2) with photocuring (PCA) and upper heated plate (UHP) accessories (TA Instruments). Samples are prepared *in situ* on 20 mm parallel geometry plates with a gap thickness $t = 250$ µm. The transparent acrylic bottom plate transmits filtered ($\lambda = 405$ nm) light from a source (Omnicure Series 1500, Lumen dynamics) into the sample. The power density is set such that the illumination is $J_e$ =0, 5, and 14 mW·cm$^{-2}$ as measured by a Model 222, G&R Labs Inc. Radiometer (405 nm probe). For the first stage, a dynamic time sweep test was performed using a constant strain ($\gamma = 1\%$) and frequency ($\varpi =1$ Hz) at 25°C while controlling photoexposure. In the second stage, the UHP ramps the temperature to 80°C at the rate of 10°C·min$^{-1}$ and holds for 1 h at a constant strain ($\gamma = 0.1\%$) and frequency of ($\varpi =0.1$ Hz). In the third stage, the UHP increases the temperature to 120°C at a rate of 10°C·min$^{-1}$ for a hold time of 10 h while maintaining the previous strain ($\gamma = 0.1\%$) and frequency ($\varpi =0.1$ Hz). Data collection occurs at a frequency of 1 Hz. The initial or "dark" viscosity is calculated by averaging the viscosity over the first 10 seconds prior to illumination.

*Fourier Transform Infrared Spectroscopy (FTIR)*: We infer polymerization kinetics using a Fourier Transform Infrared (FTIR) Spectrometer (Bruker Invenio R) with Variable Angle (ZnSe crystal 45° flate plate, 30° angle of incidence, Pike Technologies) and Heated ATR Stage (Pike Technologies) that tracks the real-time IR peaks in reflection mode. Series scans were recorded with spectra taken at the rate of 1 second per scan for photoirradiation (1 scan with a resolution 4 cm$^{-1}$) and 4 second per scan for elevated temperature measurements (16 scans with resolution 4 cm$^{-1}$). Irradiation was performed using a UV light source (Omnicure Series 1500, Lumen dynamics) with 405 nm bandgap filter ($J_e$ = 5 or 14.5 mW·cm$^{-2}$) attached to a waveguide and collimator. We determine the Conversions of thiol (2500 to 2600 cm$^{-1}$), ene (3100 to 3200 cm$^{-1}$) and epoxide (900 to 930 cm$^{-1}$) functional groups by monitoring the disappearance of the corresponding IR peaks (**Figure S2**).

*Tensile Tests:* We fabricated tensile sample with a commercial desktop SLA printer (Ember by Autodesk) using a blue-light LED projector ($\lambda$ = 405 nm) for different exposure times (t = 0, 3.5, 4.5, 6.5 and 15s). Samples were cured between two glass plates coated with a thin Teflon sheet. We measured a light intensity of $J_e$ = 30.3 mW·cm$^{-2}$ immediately above the bottom Teflon coated plate using a Model 222, 405 nm probe, G&R Labs Inc. radiometer. The final tensile samples possessed a geometry of ~10mm x 1mm x15 mm (width x thickness x height). An Instron Universal Testing System (Model 5943) with pneumatic clamps ($\Delta P$=50 psi) pulled these samples at a rate of 75mm min$^{-1}$ while a video extensometer optically tracked strain between two manually placed ink dots in the gage region of the coupon.

*Nanoindentation Measurements:* The nanoindentation films were prepared by casting in glass molds. Samples were cured by a UV light source with 405 nm bandgap filter ($J_e$ = 30 mW·cm$^{-2}$) through a photomask for 10s. The nanoindentation measurements were conducted using a MTS nanoindetation XP system equipped with a 50-$\mu m$-diameter diamond Berkovich tip (Micro Star Technologies, TB 22993). We apply a maximum 80% precent to unload with loading and unloading strain rate of 0.1 s$^{-1}$. The resulting load vs. indentation depth curves provide data specific to the location's mechanical properties. We calculate the reduced modulus ($E_r$) was calculated from the unloading the curve using the Oliver and Pharr model[69].

*Dynamic Mechanical Analysis:* We prepared the DMA samples by injecting resins between two glass sides with 0.5 mm thickness spacers irradiating by a UV light source with 405 nm bandgap filter. The samples were exposed for different exposure times (t = 0, 3.5, 4.5, 6.5 and 15s). Mechanical properties of the samples were characterized with DMA Q850 (TA Instruments). Specimens were measured in multifrequency strain mode by applying a sinusoidal stress of 1Hz frequency and 0.1% strain with the temperature ramping at 3 °C min$^{-1}$. The sample dimensions were 0.5×5×25 mm (*t*×*w*×*h*).

*Differential Scanning Calorimetry:* All DSC experiments were conducted using a DSC 2500 (TA instruments) at a heating rate of 5 °C min$^{-1}$. The glass transition temperatures were determined as the inflection temperatures in the DSC curves.

*Surface Energy Measurements:* Double sessile drop tests were conducted using a Kruss Mobile Surface Analyzer (MSA). In this experiment, the MSA deposits drops (V = 0.5 µL) of two probe liquids (water and diiodomethane) simultaneously while an on-board camera records and measures the contact angles with the substrate (N≥7). From this data (**Figure S7**), the instrument calculates the total surface energy as well as the polar and dispersive components using the Owens, Wendt, Rabel and Kaelble (OWRK) method.

*3D Printing*: All objects are printed on a consumer grade desktop SLA printer (Ember by *Autodesk*) using a blue-light LED projector (λ = 405 nm, $J_e$ ~ 30.3 mW·cm$^{-2}$). Each layer is exposed for an equivalent period (*t* = 10s). Autodesk Print Studio software sliced the CAD models into discrete 0.250 mm layers and created a corresponding black and white image stack. These images were greyscaled using image software (Microsoft Paint ©) to yield our target photodosage per pixel.

## Acknowledgements

We appreciate the guidance from Tyler Christensen when integrating our braille display with pneumatic valving. We thank Joseph Aase and Sean Braanten for their support fabricating test fixtures and documenting the operation of our devices. Portions of this work was performed under

## Author Contributions

T.W., S.H., S.A., J.S., P.P., M.S., and Y.M. conceived and planned the experiments. T.W., S.H., S.A., J.S., and P.P. carried out the experiments. T.W., S.H., J.S., and S.H. contributed to the interpretation off the results. S.H. and T.W. took the lead in writing the manuscript. All authors provided critical feedback and helped revise the manuscript.

## Competing Interests

The authors declare the following competing interests: Facebook Reality Labs provided equipment, supplies, salaries, and other expenses for some of the authors, and Facebook has disclosed portions of this work to the United States Patent and Trademark Office as Provisional Patent US17062355: Ternary Orthogonal Photochemistries.

## Data Availability

The authors declare that the data supporting the findings of this study are available within the paper and its Supplementary Information files or from the corresponding authors upon reasonable request

## Code Availability

The authors declare that no custom computer code or algorithm were used to generate results that are reported in the paper

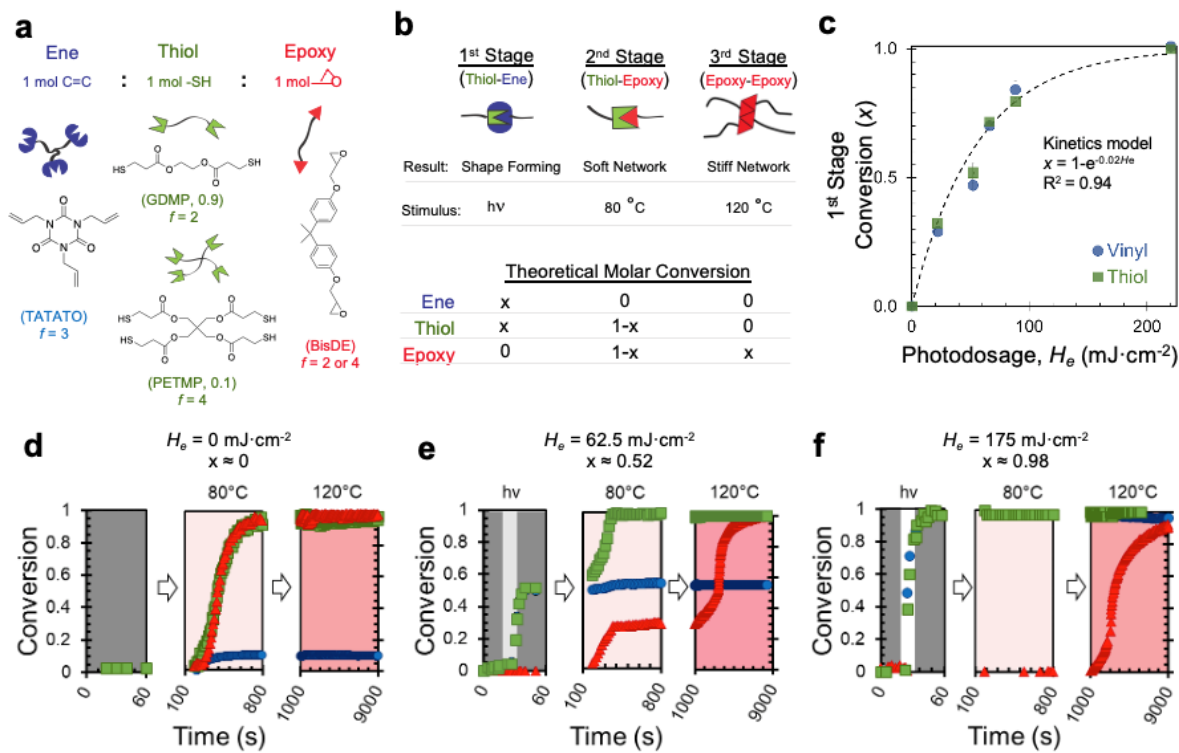

**Figure 1: Ternary Sequential Reaction Scheme**. a) The polymer precursors of our resin blended in a 1:1:1 ene:thiol:epoxy molar ratio. b) The stimuli, result, and molar conversion for each reaction stage. c) The functional group conversion ($x$) of thiol and ene groups during the first stage as a function of applied photodosage, $H_e$. Error bars represent standard deviation (N=3). d-f) The functional group conversion of ene, thiol, and epoxide groups for each reactive stage as a function of initial applied photodosage, $H_e$.

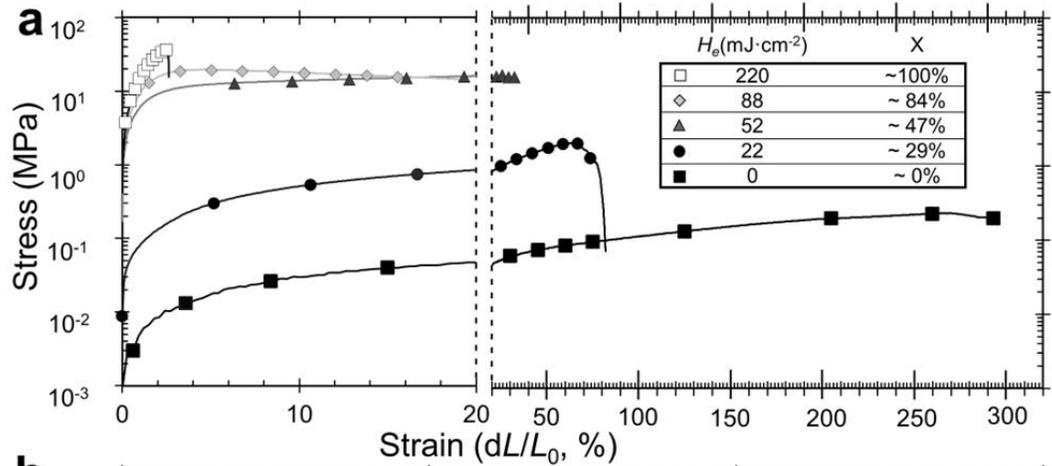

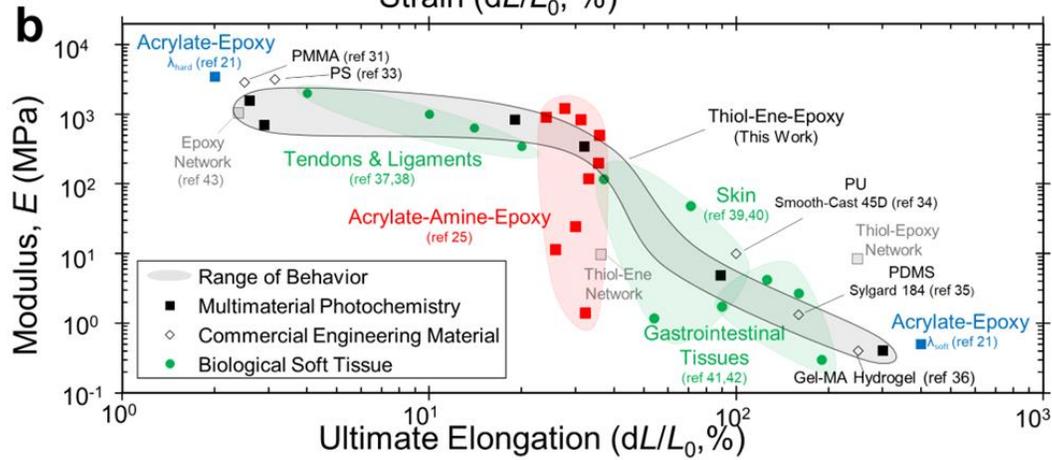

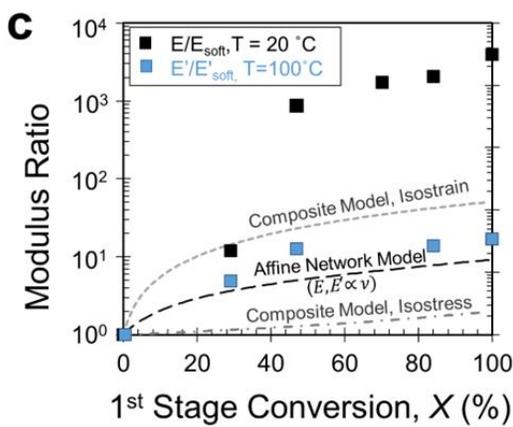
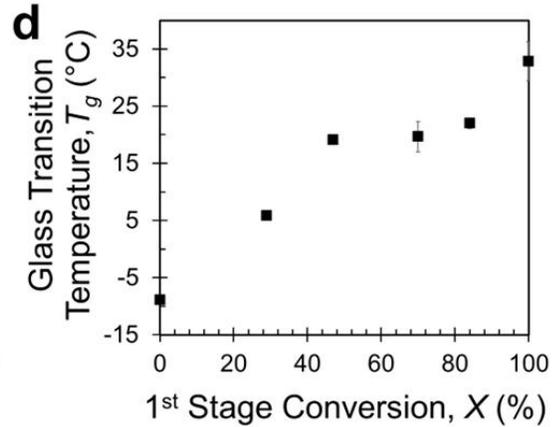

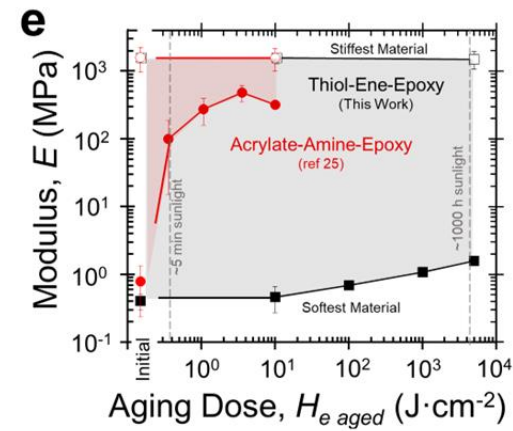
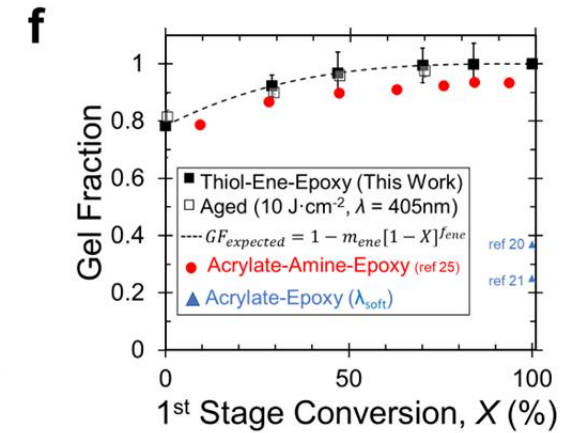

**Figure 2: Stable Multimaterial Properties**. a) The average stress-strain curves for our fully processed multimaterial chemistry as a function of initial photodosage or 1$^{st}$ stage conversion (N $\geq$ 7). Please note the log-linear scaling and the break in the x-axis. b) Comparison of available mechanical properties between commercial engineering materials, biological tissues, and multimaterial photochemistries. Data for other materials obtained from ref. 21, 25, 31-43. c) The modulus ratio of our materials at room temperature and 100 ˚C compared to that predicted by composite models and the affine network model. d) The glass transition temperature as a function of photoexposure (error bars represent standard deviation, N $\geq$ 3). e) Average Young's Modulus of the stiffest and softest material as a function of post process photoirradiation ($\lambda$ = 405 nm, error bars represent standard deviation, N $\geq$ 7). Shaded regions represent available intermediate materials f) Gel fraction as a function of 1$^{st}$ stage conversion for representative multimaterial photochemistries. Our system exhibits a high average gel fraction (error bars represent standard deviation, N $\geq$ 7) that remains stable after UV aging ($H_{e,\ aged}$ = 10 J cm$^{-2}$, $\lambda$ = 405 nm) which suggests geometric stability and no further polymerization after processing. Data for other materials obtained from ref. 20,21,25.

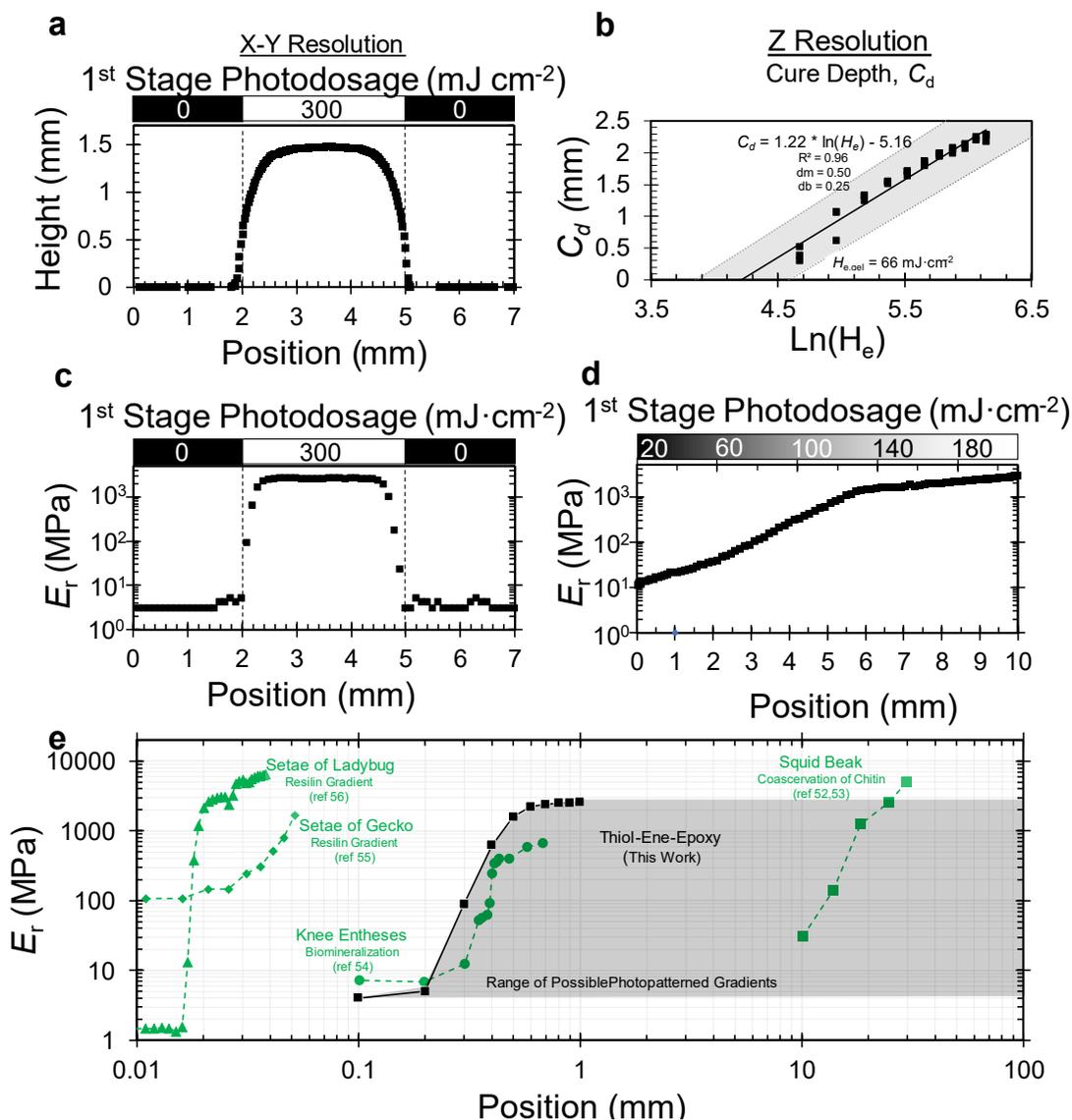

**Figure 3: Shape and Mechanical Property Resolution in Photocured Samples. a)** The average surface profile that results from a 3mm wide photoexposure of 300 mJ·cm$^{-2}$ followed by immediate removal of unreacted precursors (N =7). The x-y resolution of this gelled material is within 4% of the target photopattern at the layer heights common to 3D printing (100 μm). **b)** Average cure depth, $C_d$, as a function of applied photoexposure, $H_e$ (N = 3). Shaded regions capture the uncertainty in the best fit equation. Maximum z-resolution occurs when the $C_d$ matches the target layer height. **c)** Fully processing a multimaterial film based on a single 3mm wide photoexposure of 300 mJ·cm$^{-2}$, we measure the location dependent pressure-displacement curves via nanoindentation and extract the reduced modulus, $E_r$, from the unloading curve. Despite an abrupt transition in applied photodosage, we observe a continuous mechanical gradient. **d)** A grayscaled photopattern enables programming shallower mechanical gradients as confirmed by nanoindentation. **e)** Comparison of our photopatterned gradients to those that exist in nature. We plot the data sets on the decade that best describes their length scale. Black squares represent the transition measured in **Figure 3c** and shaded regions represent the range of possible shallower gradients obtained via grayscale patterning. Data for other materials obtained from ref. 52-56.

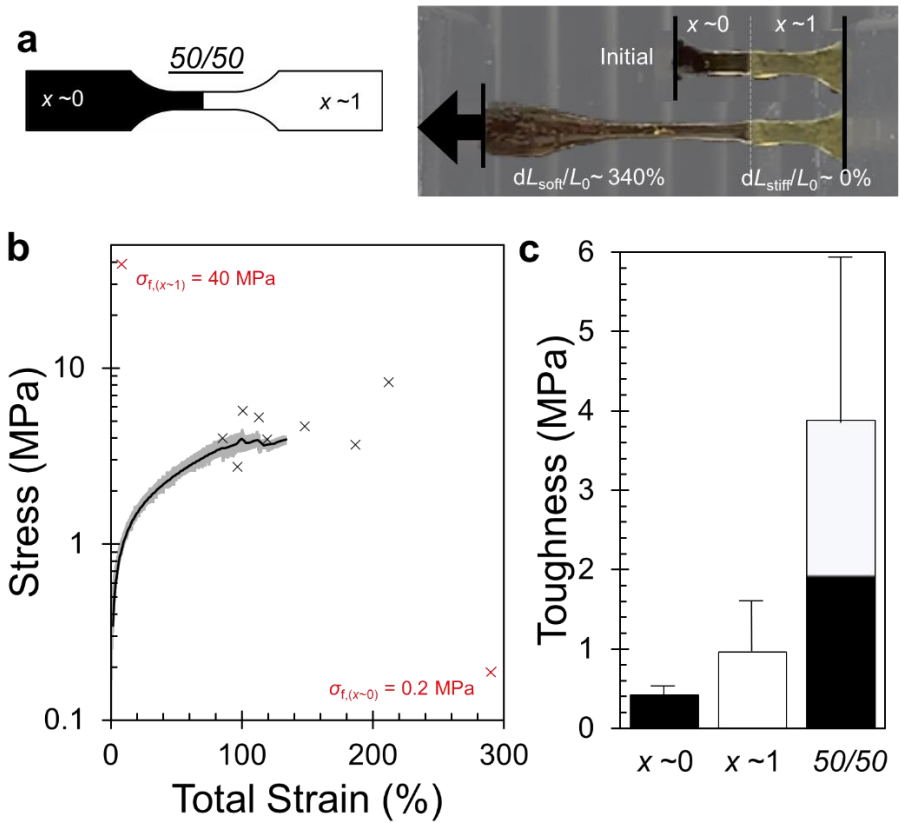

**Figure 4: Strain Localization at Tough Multimaterial Interfaces.** a) Photopattern to create a dog bone comprised of our softest ($x \sim 0$, $H_e = 0$ J cm$^{-2}$) and stiffest material ($x \sim 1$, $H_e \sim 0.9$ J cm$^{-2}$) with representative images from a tensile test showcasing extreme strain localization to the soft region (d$L/L_0 \sim 340\%$). b) The average stress-strain curve of these multimaterial dogbones under applied load (N = 8). Shaded regions correspond to max and min stresses at each strain value, failure points are denoted by "x," and the average failure points of the pure components are shown as a red "x" for comparison, see Figure 2a for full data. c) These multimaterial dogbones exhibit a statistically significant (unpaired, two tailed t-test, α =0.05) enhancement in tensile toughness (~10-fold) when compared to the pure components.

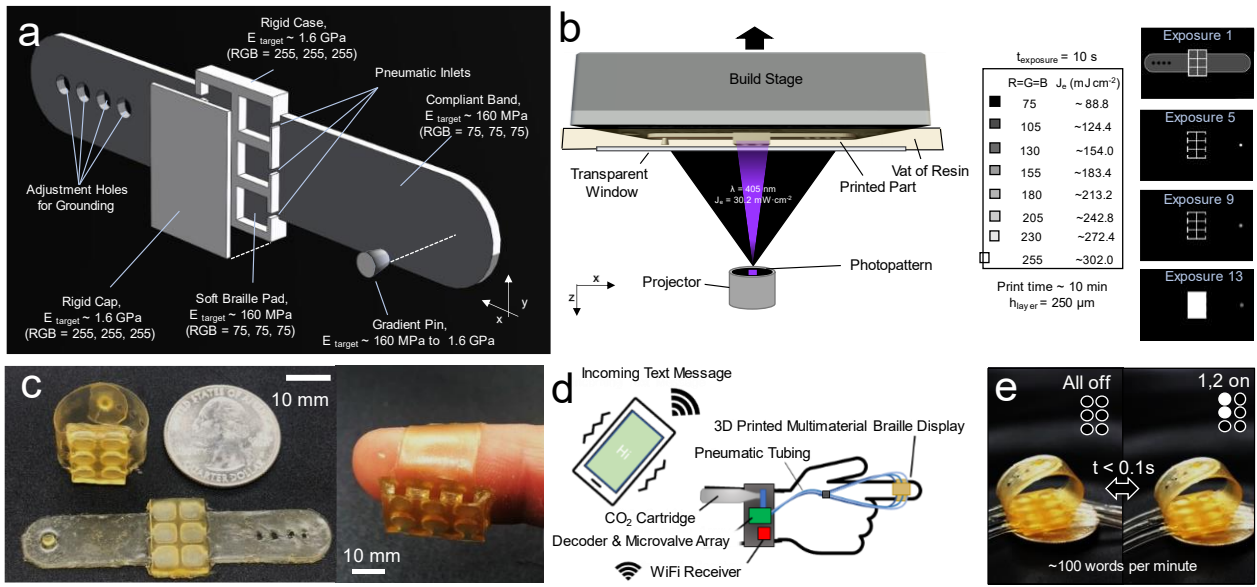

**Figure 5: Multimaterial 3D Printed Soft, Wearable Braille Display.** a) Detailed view of the design for a soft, wearable braille display. The design is greyscaled to provide local control over modulus. b) Overview of bottom-up projection-based SLA 3D printing process. We expose each 250 μm layer for 10 s and the pixel is greyscaled to achieve the target layer photodosage with the representative photopatterns for corresponding layers. c) As printed braille displays showing compliant bending of the band and conformal fit around a finger. d) Proposed untethered operation of a wearable braille display for mobile interaction. e) High frequency operation of a braille display coupled to pneumatic tubing and valving with refresh rates of ~100 words per minute. See Supplemental Video 1 for a demonstration.

**One-Pot Printing of Robust Multimaterial Gradients**

*Sijia Huang\*, Steven Adelmund, Pradip S. Pichumani, Johanna J. Schwartz, Yiğit Mengüç, Maxim Shusteff, Thomas J. Wallin\**

**Video S1: Multimaterial 3D Printed Braille Display During Operation**

1. **Composite Modelling**

One method to attempt to describe the multimodulus behavior of the thiol-ene-epoxy system is to consider the final material a composite of the three pure individual networks (i.e. thiol-ene, thiol-epoxy, and epoxy). Starting form Hooke's Law (Eq. S1), we can derive the modulus of a three phase composite ($E_{tot}$) for two situations: 1) when all domains experience the same stress (isostress, $\sigma_{tot} = \sigma_1 = \sigma_2 = \sigma_3$) but the total strain is a volume fraction weighted sum of the individual strains, or 2) when all domains experience the same strain (isostrain, $\varepsilon_{tot} = \varepsilon_1 = \varepsilon_2 = \varepsilon_3$) and the total stress is a volume fraction weighted sum of the individual stresses.

Isostress Condition:

$$E = \frac{\sigma}{\varepsilon} \quad \text{(Eq. S1)}$$

$$\sigma_{tot} = \sigma_1 = \sigma_2 = \sigma_3 \quad \text{(Eq. S1.2)}$$

$$\varepsilon_{tot} = v_1 \varepsilon_1 + v_2 \varepsilon_2 + v_3 \varepsilon_3 \quad \text{(Eq. S1.4)}$$

$$\varepsilon_{tot} = v_1 \frac{\sigma_1}{E_1} + v_2 \frac{\sigma_2}{E_2} + v_3 \frac{\sigma_3}{E_3} \quad \text{(Eq. S1.6)}$$

$$\varepsilon_{tot} = \frac{v_1 E_2 E_3 \sigma_{tot} + v_2 \sigma_{tot} E_1 E_3 + v_3 \sigma_{tot} E_1 E_2}{E_1 E_2 E_3} \quad \text{(Eq. 1.8)}$$

$$E_{tot} = \frac{\sigma_{tot}}{\varepsilon_{tot}} = \frac{E_1 E_2 E_3}{v_1 E_2 E_3 + v_2 E_1 E_3 + v_3 E_1 E_2} \quad \text{(Eq. S1.10)}$$

Isostrain Condition:

$$E = \frac{\sigma}{\varepsilon} \quad \text{(Eq. S1)}$$

$$\varepsilon_{tot} = \varepsilon_1 = \varepsilon_2 = \varepsilon_3 \quad \text{(Eq. S1.3)}$$

$$\sigma_{tot} = v_1 \sigma_1 + v_2 \sigma_2 + v_3 \sigma_3 \quad \text{(Eq. S1.5)}$$

$$\sigma_{tot} = v_1 E_1 \varepsilon_1 + v_2 E_1 \varepsilon_1 + v_3 E_1 \varepsilon_1 \quad \text{(Eq. S1.7)}$$

$$\sigma_{tot} = (v_1 E_1 + v_2 E_1 + v_3 E_3) \varepsilon_{tot} \quad \text{(Eq. S1.9)}$$

$$E_{tot} = \frac{\sigma_{tot}}{\varepsilon_{tot}} = (v_1 E_1 + v_2 E_1 + v_3 E_3) \quad \text{(Eq. S1.11)}$$

Where $v_i$ is the volume fraction of that phase in the composite and $E_i$ is the modulus of that phase.

We fabricated samples (N≥4) comprised of only the thiol-ene or thiol-epoxy components and their corresponding intiators and measured the resulting mechanical properties ($E_{thiol-ene}$ = 9.9 MPa, $E_{thiol-epoxy}$ = 8.5 MPa) and density ($\rho_{thiol-ene}$ = 1.3 g mL$^{-1}$, $\rho_{thiol-epoxy}$ = 1.3 g mL$^{-1}$). For the pure epoxy material, we utilized data from a previous study ($E_{epoxy}$ = 1000 MPa, $\rho_{epoxy}$ = 1.2 g mL$^{-1}$)[1]. We can also estimate the mass fraction of each phase or polymer network in the composite based on the first stage photoconversion (*X*) as shown in Figure 1B and the mass fraction of the monomers in the initial composition.

$$m_{thiol-ene\ network} = (m_{thiol} + m_{ene})X \quad \text{(Eq. S1.12)}$$
$$m_{thiol-epoxy\ network} = (m_{thiol} + m_{epoxy})(1 - X) \quad \text{(Eq. S1.13)}$$
$$m_{epoxy\ network} = (m_{epoxy})X \quad \text{(Eq. S1.14)}$$

Where $m_i$ is the mass fraction of each monomer in the original resin (see Table S1). To convert this mass fraction to a volume fraction, we consider the densities of the polymer networks.

$$v_{thiol-ene} = \frac{m_{thiol-ene}\,\rho_{thiol-ene}^{-1}}{m_{thiol-ene}\,\rho_{thiol-ene}^{-1} + m_{thiol-epoxy}\,\rho_{thiol-epoxy}^{-1} + m_{epoxy}\,\rho_{epoxy}^{-1}} \quad \text{(Eq. S1.15)}$$

$$v_{thiol-epoxy} = \frac{m_{thiol-epoxy}\,\rho_{thiol-epoxy}^{-1}}{m_{thiol-ene}\,\rho_{thiol-ene}^{-1} + m_{thiol-epoxy}\,\rho_{thiol-epoxy}^{-1} + m_{epoxy}\,\rho_{epoxy}^{-1}} \quad \text{(Eq. S1.16)}$$

$$v_{epoxy} = \frac{m_{epoxy}\,\rho_{epoxy}^{-1}}{m_{thiol-ene}\,\rho_{thiol-ene}^{-1} + m_{thiol-epoxy}\,\rho_{thiol-epoxy}^{-1} + m_{epoxy}\,\rho_{epoxy}^{-1}} \quad \text{(Eq. S1.17)}$$

We can now solve for the composite modulus $E_{tot}$ as a function of *X* for both the isostress and isostrain conditions. By dividing *E(X)*, by the softest modulus, or *E(X=0)*, we can generate the predicted modulus ratio as shown in Figure 2c. We find that the isostress model predicts a modulus ratio ($E_{stiff}/E_{soft}$) of only 2.0, while the isostrain model predicts a ratio of 51.0. Both of these values are orders of magnitude below our observed room temperature ratio of 3,800. In practice, the isostress or isostrain models only accurately describe limited situations like laminated structures loaded along the principal axes. Instead, the isostress and isostrain models typically bound the observed behavior. However, our observed material exhibits a modulus ratio that far exceeds those predicted by such models. Such estimations assumes that the three polymer networks do not intermix at small lengthscales, but form discrete domains. As we expect thiol, ene, and epoxy subunits to all add to the same polymer network at intermediate conversions (0 < *X* < 1), it is unsurprising that the material does not behave as a conventional composite. Other mechanisms must be contributing to the mechanical performance.

## 2. Polymer Structure-Property Relationship:

Our material chemistry uniquely exhibits a large range of stiffness ([$E_{stiff}/E_{soft}$] > 3,800) from a single resin composition. We attribute this performance to the synergetic effect of variable crosslink density and local chain rigidity. Flory's affine network model[2] relates the rubber modulus (*E)* to the molar crosslink density, ν, as shown in Equation S2:

$$E \propto RT\nu \quad \text{Equation S2}$$

Where R is the gas constant and T is the absolute temperature.

We define a crosslinker as a species that forms more than two bonds with the polymer network. As such, we can calculate the number of crosslinks formed by multiplying the number of crosslinking reactions (i.e., any bonds in excess of the first two) for a given reactive scenario by the probability of that scenario occurring. We then sum overall all possible conditions and weight according to relative mol fraction (*m*) of that species in the resin.

We utilize only one ene species in our resin ($m_{TATATO} = 1$) and it is trifunctional ($f_{TATATO} = 3$). Thus, a single, ene-based crosslink forms when a TATATO monomer fully reacts (Equation S2.2). For our specific system, the number of thiol-based crosslinks is simple to calculate as we assume all thiols react in a stepwise addition manner during the first two stages. Only two of the thiol groups on our tetra-functional PETMP species act as crosslinkers (the other two thiol groups on PETMP and both thiols on GDMP merely act as chain extenders). Thus, twice the thiol mol fraction of PETMP ($m_{PETMP} = 0.1$) form crosslinks. However, this construction, when combined with Equation S2.2, "double counts" crosslinks formed by the ene and PETMP. If we assume equal reactivity between GDMP and PETMP, we can remove the equivalent mol fraction of ene crosslinks (see Equation S2.3). Calculating the number of epoxide-based crosslinks requires consideration of whether these groups react during the second or third stage. As shown in Figure S1, an epoxide group that participates in step addition has a functionality of one (adds to one thiol) but exhibits a functionality of two (adds to two other epoxides) when participating in chain growth. Once fully reacted, there are three possibilities for our diepoxy monomer: (1) both epoxides participate in thiol-epoxy addition ($f = 2$, no crosslinks formed); one epoxide adds to a thiol, one epoxide participates in chain growth ($f = 3$, one crosslink formed), or both epoxides participate in chain growth ($f = 4$, two crosslinks formed). These crosslinking scenarios are captured mathematically in Equation S2.4. We can use Equations S2.2-S2.4 to calculate the ratio of crosslink density at a given 1$^{st}$ stage conversion, $\nu(x)$, to the crosslink density found in our softest fully processed material ($\nu_{x\,=\,0}$) as shown in Equation S2.5.

$$\nu_{ene} = (1)\, m_{TATATO}(x)^{f_{TATATO}} = (x)^3 \qquad \text{(Eq. S2.2)}$$

$$\nu_{thiol} = 2m_{PETMP} - m_{PETMP}\eta_{ene} = 0.2 - 0.2(x)^3 \qquad \text{(Eq. S2.3)}$$

$$\nu_{epoxy} = 0\,(1-x)^2 + 1(1-x)(x) + 2(x) = x^2 + x \qquad \text{(Eq. S2.4)}$$

$$\frac{E}{E_{soft}} \propto \frac{\nu(x)}{\nu_{x\,=0}} = \frac{0.8x^3 + x^2 + x + 0.2}{0.2} \qquad \text{(Eq. S2.5)}$$

Unfortunately, this model only predicts a maximum of ~15x change in chemical crosslink density, and by extension modulus, between our soft and stiff regions ($x \sim 0$ to $x \sim 1$). This result is orders of magnitude below experimental observation. Thus, another phenomenon must be contributing to our measured moduli.

The affine network theory (above) assumes the polymer chemistry is uniform throughout the network and therefore each chain exhibits similar molecular mobility. In actuality, the chain mobility varies according to the main-chain and side group chemistry. Covalent bonds along the backbone possess characteristic bond lengths, energies, and angles which, along with the steric hindrances of any substituents, determine how freely polymer chains can rotate and elongate. Segments that cannot rotate easily form stiff domains. For example, the short BisDE (see Figure 1A) monomer is relatively rigid because of restricted rotation about the bulky phenyl rings when compared to the ethylene glycol di[3-mercaptopropionate] (GDMP) species. In GDMP, the –(C=O)-O- repeat unit is particularly flexible due to the $sp^2$ hybridization of the backbone—whereas normal aliphatic polymers possess a sidegroup above and below the chain, the oxygen linkage has no side groups (only lone pairs) and the carbon is only affixed to a single side group (=O). Thus, there is less steric hindrance to rotation about these subunits. As mentioned above (Figure S1), the chemical subunits we add to our polymer backbone vary by reaction mechanism (e.g. thiol-ene and thiol-epoxy reactions incorporate the length of the flexible di-ethylene glycol chains into the backbone whereas epoxy-epoxy reactions add short –[C-O]– repeat units with bulky phenyls). Thus, we expect this distinction to exert a profound effect on the chain's "rigidity" to further contribute to the stiff moduli observed at large $x$.

Beyond covalent bonds between chains, other interactions may function a "effective crosslinks." For example, at high 1st stage conversions, ($x\rightarrow1$), the first stage forms a thiol-ene network leaving no thiols for grafting epoxy in the second stage. In these instances, the epoxy-epoxy homopolymerization forms an almost independent, secondary polymer network. Any resulting entanglements between these two interpenetrating networks act as physical crosslinks to add to the chemical crosslink density. Alternatively, the strength of some intermolecular interactions (hydrogen bonding, pi-stacking) is entirely dependent on the proximity of molecular groups. It is possible that BisDE homopolymerization results in a favorable arrangement of aromatic rings to promotes strong pi-stacking and further stiffen the network at high conversions ($x$). We attribute the deviations that remain between the affine network model and our observed high temperature (i.e. T $\gg$ T$_g$) modulus ratio to such non-chemical effective crosslinks.

### 3. Print Parameters and Design Considerations for Multimaterial 3D Printing

To 3D print our photochemistry, we need to identify and use the proper process parameters. To aid in this attempt, we utilize an open source stereolithography (SLA) 3D Printer (*Autodesk, Inc* Ember) that permits modification of layer height, photoexposure time, individual photopatterns, etc. In this monochromatic ($\lambda$ = 405 nm) projection-based system, the entire photopattern is illuminated for the same length of time which requires utilizing greyscale in order to achieve variable pixel dosages. Equation S3 calculates the greyscale value ($Y$) for any arbitrary color in the RGB model, but for simplicity, we only use colors along the white-black spectrum (i.e, $R=G=B$). We then create an image stack of full color exposures at varying hues from white-black (0 < $Y= R=G=B$ < 255) while measuring photoirradiative power immediately above the build window (Model 222, 405 nm probe, *G&R Labs Inc.*). As shown in Figure S9a, a simple linear fit describes the data very well (R > 0.99). From the FTIR data, we note that the 1st stage conversion saturates at $H_e$ = 220 mJ·cm$^{-2}$. Thus, we select an exposure time of $t_{exposure}$= 10s such that our full-white dose ($Y$ = 255, $J_e$ ~ 29 mW·cm$^{-2}$) is sufficient for full 1st stage conversion. Since only integral values of greyscale are possible (i.e. $\Delta Y_{min}$ = 1), a longer exposure time is necessary to obtain less than 1.13 mJ·cm$^{-2}$ changes in photoexposure between pixel colors. However, we note that this extends the total print time.

$$Y = 0.2126R + 0.7152G + 0.0722B \qquad \text{(Eq. S3)}$$
$$Y = R = G = B, for\ R = G = B$$

Understanding the cure depth as a function of photoexposure is key to realizing high resolution printing. We solve this relation experimentally by exposing the resin to varying photoirradiation dosages and measuring the corresponding height of the cured object. To collect all data in a single experiment, we create a photopattern (Figure S9b) containing an array of greyscaled circles. Once printed, we measure the heights of the corresponding cylinders using 3D profilometry (Keyence VK-3200 G2 series microscope). We then infer ($C_d$) as the max height of these cylinders and plot these values as a function of the natural log of photoexposure. However, there is often some variance in pixel intensity such that we recommend a photodosage 5-10% greater than that required for the layer height according to these measurements. As an aside, from the linear fit in **Figure 3b**, we calculate a critical gel dosage of 67.5 mJ cm$^{-2}$. According to the best fit of Figure 1c, this dose yields a conversion, $x$~0.7, which is slightly above the gel threshold ($x$~0.65) predicted by Flory Stockmayer Theory (A:B = 1:1, $f_{ene}$ ~ 3, $f_{thiol}$ ~ 2.2).

When targeting a specific modulus, we need to determine a desired photoexposure. Unfortunately, while we were able to develop a simple reaction model that connects photoexposure to 1st stage conversion (Equation 1), we were unable to theoretically predict modulus from conversion (see **Figure 2d**). As such, we rely on an empirical fit (Figure S9e) to determine $E(x)$. For applications where $E$ is critical, we recommend sampling more data points to improve interpolation of this relationship. This fit, along with that of Equation 1, enables extrapolating target $H_e$ for a given $E$. However, situations may arise for low target $E$ where $H_e$ is below $H_{e,gel}$. In these situations, the material does not solidify. While this isn't an issue for 2D fabrication, such areas will not survive the build translations in conventional 3D printing. In these cases, another option is to "encapsulate" these underexposed layers in a shell of solid material, a process known as "cavity vat polymerization"[3]. Such shells preserve the volume partially cured resin until thermal curing forms the solid.

In SLA, the object is built through a series of successive 2D exposures. As mentioned previously, the light can penetrate the resin and even into previously cured layers. While not an issue for conventional, single-property materials, such "cure through" can increase the actual applied photoexposure ($H_e$) and conversion ($x$) to alter the material properties in our material. We can utilize the penetration depth ($D_p$) to precisely calculate the cumulative photoexposure and "underexpose" the printed layer to accommodate cure through. Additionally, we could change the layer height of our print design, but this creates a tradeoff between numerous small layers (short exposure times) and a few larger layers (longer exposure times). We found a layer height of 250 microns to be a suitable balance between these considerations. The inclusion of dyes can also alter the attenuation behavior of a resin to aid in the optimization of penetration depth, layer height, and cumulative photodosage (see Figure S9).

The design and orientation of a part is also critical to obtaining a successful print. First, the part needs to survive the translations during printing. For bottom-up SLA, the interplay between adhesion to the build stage/previously printed layers and release from the build window requires avoiding large changes in cross sectional area along the print (z) direction. Further, to avoid cure through issues, the part should be oriented to minimize the subsequent illumination of soft voxels in the image stack. For a given point in x,y space, stiff materials should be printed in earlier layers than soft materials. Thus, we designed our braille display such that the compliant band is directly on the build stage and for robustness during shearing, we patterned a thin, white (stiffer) line around the edge.

Another common challenge when SLA printing fluidic elastomer actuators, like our braille dots, results from incomplete draining of enclosed cavities. The suction forces that result from unvented features prevents resin recirculation. This entrained liquid can slowly polymerize to gel and result in loss of feature resolution. To aid in resin flow, we enlarged the size of our inlets, braille pads, and added to the height of our rigid case. Still, the removal of liquid resin from these channels post-print required tedious cleaning with a cotton swab and isopropyl alcohol.

For situations where the above recommendations are difficult to implement in a single print, we recommend a combined print-assemble-bond approach. Prior to the thermal treatment, the as-printed objects still possess the precursors of the 2$^{nd}$ and 3$^{rd}$ stages. These reactive groups are sufficient to provide cohesive bonding across printed objects ($0 < x < 1$) that are in contact during final curing. While we were still able to directly print working braille displays, by moving the "rigid cap" from the top of the "rigid case" and on to build stage, we were able to mitigate issues with insufficient drainage and cure through. Cleaning of internal features also became much easier, and we were able to assemble the cap on top of the device prior to heating. After full processing, the device held pressure and operated similarly to its directly printed counterpart.

4. **Alkene Homopolymerization after Processing**

All polymers are susceptible to aging, particularly under high intensity, deep UV exposure. Yet, multimaterial chemistries based on underpolymerization of reactive groups suffer drastic changings in performance under even modest conditions (300 mJ cm$^{-2}$, $\lambda$ = 405 nm). By comparison, our ternary thiol-ene-epoxy system does not exhibit a change in modulus or gel fraction with aging despite possessing unreacted ene groups. This result is surprising as alkene homopolymerization is not only possible, but a common industrial process.

One factor that may contribute to the stability in system relates to the nature of our chosen ene species (TATATAO). Not all (C=C) groups exhibit similar reactivities. Many of the multimaterial photochemistries rely on acrylates which usually more readily homopolymerize than our corresponding allyl groups. While idiosyncrasies exist for specific molecules, in general the stability of the α C-H bond determines the rate of chain transfer (and therefore polymerization rate) of the ene. Accordingly, the conjugated triazine trione structure may impart stability to the radical intermediates of TATATO, particularly when compared to those of acrylates, that slow the kinetics. We want to emphasize that we overcome the reduced reactivity of our alkene species by relying on thiol-ene addition during photocrosslinking and not chain growth homopolymerization, like other multimaterial chemistries. Otherwise, we would expect a drastically slower print speed and worse resolution (more time for diffusion) for a chemistry based solely on TATATO homopolymerization.

Further, numerous observations support our stability argument. First, some amount of reaction may occur without altering the observed mechanical properties. As mentioned above, TATATO only contributes a crosslink when all three ene groups are reacted. Thus, a single reaction may not create an additional of a crosslink and result in a measurable increase in observed modulus. Still, our gel fraction as a function of conversion agrees with expectations of extracting the fully unreacted TATATO molecule. Even a minor degree of homopolymerization during aging should result in an observed increase in gel fraction. Instead, our gel fraction is consistent. Hence, we propose that for our fully processed materials the remaining TATATO molecules cannot find and orient themselves in space to permit suitable homopolymerization on the time scales of our experiment.

5. **Expected Gel Fraction as a Function of 1st Stage Conversion**

The gel fraction is the mass fraction of material in the gel network. In theory, any uncrosslinked monomer, dimer, trimer, etc. can escape the gel given enough time. In our sequential chemistry, we assume that all thiol and epoxide groups react during processing; hence, extractable species will only possess pendant enes. The smallest such molecule is the completely unreacted ene monomer. Since $X$ is the first stage conversion, then the probability that a given ene group remains unreacted is $(1-X)$. However, the ene monomer contains $f$ ene groups, so we need to calculate the probability, $p$, that all these groups have reacted, or $X^f$. In our system, the next smallest molecule is two ene monomers tethered to a single thiol monomer. This trimer results when $(f-1)$ ene groups remain unreacted on each ene monomer and at least two thiols are reacted on that species. Mathematically this probability is represented as $(1-X)^{f-1}(1-X)^{f-1}(X)^2$. We can sum over ever-increasing complex structures, but since $x < 1$ and $f > 1$ these oligomers become less and less likely to exist. Thus, we ignore these higher order terms and focus on the entirely unreacted monomer.

$$GF = 1 - \sum_i m_i p_i = 1 - [m_{ene}(1-X)^f + m_{ene-thiol-ene}(1-X)^{f-1}(1-X)^{f-1}X + \ldots]$$

$$= 1 - [m_{ene}(1-X)^f] \qquad \text{Equation (2)}$$

Where $m_i$ is mass fraction in the resin of a given species and $p_i$ is the probability that that species exists.

Table S1 A sample formulation for thiol-ene-epoxy resin system.

|  | Mw (g/mol) | Weight (g) | Cost ($/g resin) |
|---|---|---|---|
| PETMP | 488 | 0.72 | 0.0096 |
| GDMP | 240 | 6.5 | 0.0981 |
| TATATO | 254 | 5 | 0.0819 |
| BisDE | 340 | 10.2 | 0.0703 |
| Speedcure 2022 | N/A | 0.28 | 0.0290 |
| LC80 | N/A | 0.48 | 0.0031 |
| PYR | N/A | 0.02 | 0.0004 |

Table S2. Young's modulus for thiol-ene-epoxy system before and after UV aging with calculations for an unpaired, two tailed t-test. Differences between initial and aged samples at the same initial photoexposure dosage are not statistically significant ($\alpha = 0.05$).

t-Test: Two-Sample Assuming Unequal Variances ($\alpha = 0.05$)

|  | 0 mW·cm$^{-2}$ | | ~34 mW·cm$^{-2}$ | | ~79 mW·cm$^{-2}$ | | ~135 mW·cm$^{-2}$ | | ~180 mW·cm$^{-2}$ | | ~338 mW·cm$^{-2}$ | |
|---|---|---|---|---|---|---|---|---|---|---|---|---|
|  | Initial | Aged | Initial | Aged | Initial | Aged | Initial | Aged | Initial | Aged | Initial | Aged |
| Average Modulus (MPa) | 0.407 | 0.442 | 4.86 | 4.98 | 347 | 282 | 701 | 652 | 838 | 877 | 1570 | 1570 |
| Variance | 0.013 | 0.045 | 6.29 | 2.51 | 14000 | 4010 | 30500 | 8800 | 12900 | 38100 | 71600 | 88400 |
| Sample Size | 7 | 7 | 8 | 8 | 7 | 8 | 8 | 7 | 8 | 8 | 7 | 8 |
| Degrees of Freedom | 9 | | 12 | | 9 | | 11 | | 11 | | 13 | |
| t Stat | -0.379 | | -0.115 | | 1.31 | | 0.686 | | -0.492 | | 0.061 | |
| P(T<=t) two-tail | 0.714 | | 0.910 | | 0.222 | | 0.507 | | 0.632 | | 0.952 | |
| t Critical two-tail | 2.26 | | 2.18 | | 2.26 | | 2.20 | | 2.20 | | 2.16 | |

1$^{st}$ Stage:
Thiol-ene reaction
Step growth polymerization

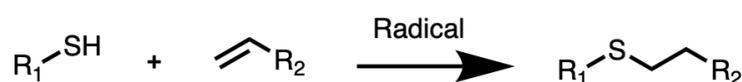

2$^{nd}$ Stage:
Thiol-epoxy reaction
Step growth polymerization

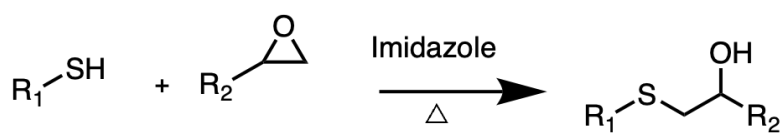

3$^{rd}$ Stage:
Epoxy homopolymerization
Chain growth polymerization

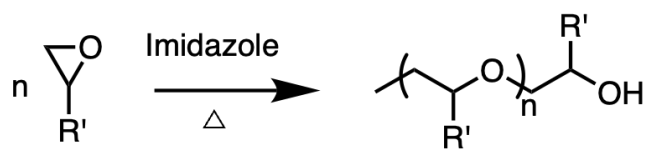

Figure S1: Reaction Schemes for One-Pot 3D Printing of Multimaterials. a) Photo-controlled thiol-ene step growth polymerization. b) Thermally initiated thiol-epoxy step growth polymerization. c) Thermally initiated epoxy chain growth homopolymerization

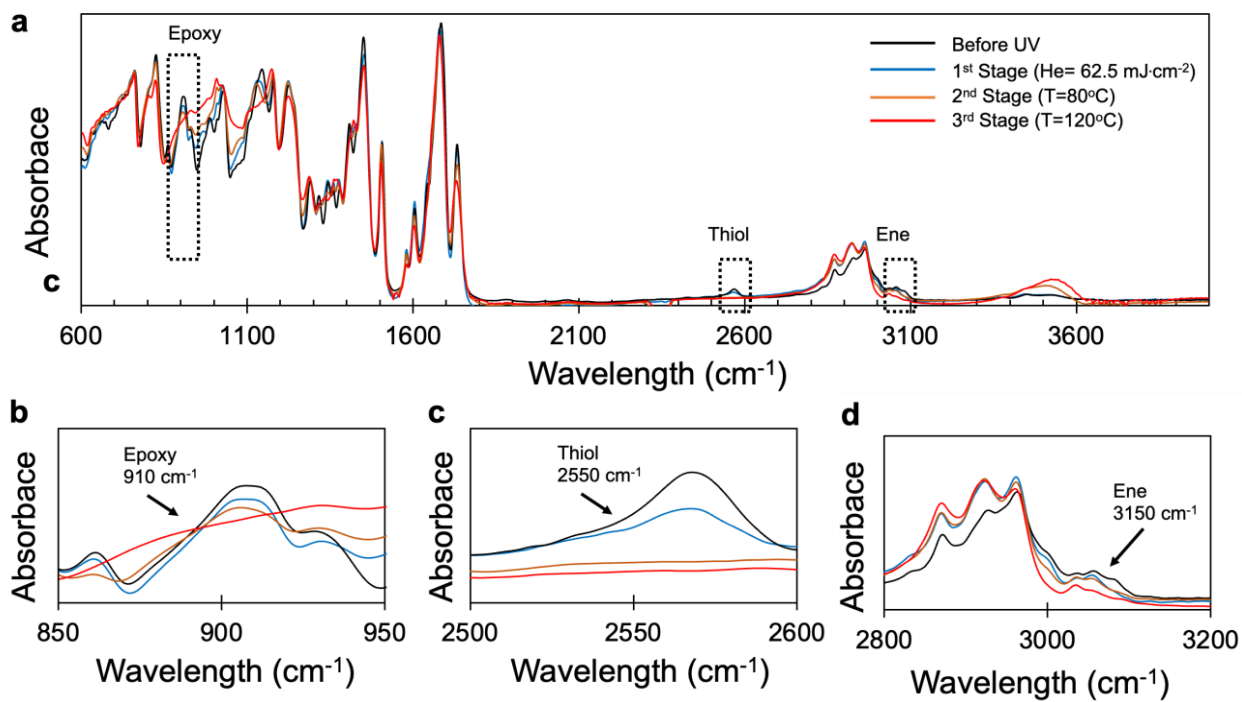

**Figure S2: FTIR Spectra of One-Pot 3D-Printing of Multimaterial.** a) FTIR spectrums of the resin before UV curing, after 1st stage, 2nd stage, and 3rd stage. b) Epoxide peak; c) Thiol peak; d) Ene peak.

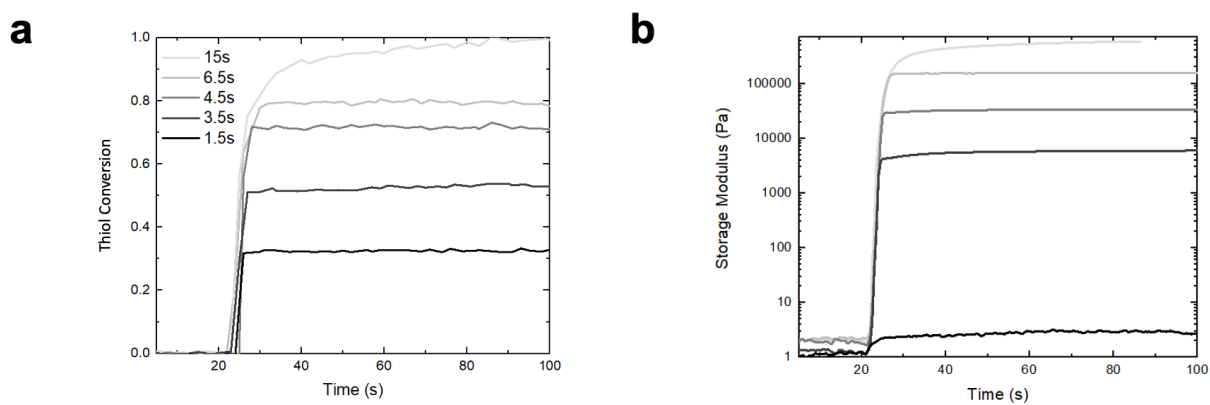

**Figure S3: FTIR Spectrums and Photorheology.** a) Thiol-ene reaction kinetic with different photodosage. (b) Change in storage modulus ($G'$) over time under different photodosage. The resin contains a 1:1 thiol : ene molar raio.

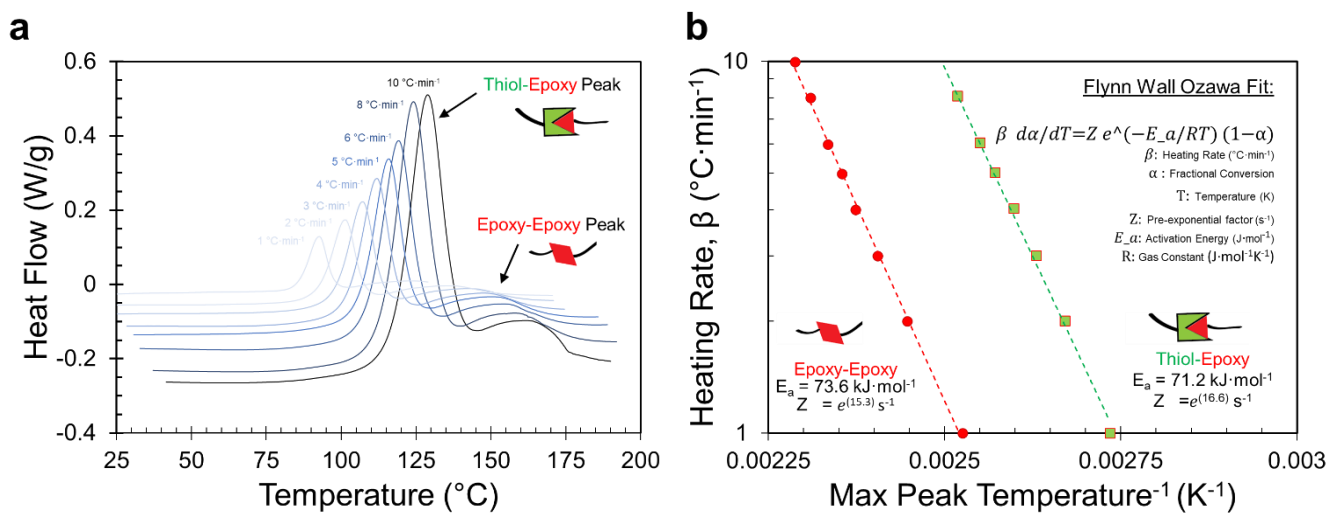

**Figure S4: Differential Scanning Calorimetry of Thiol-epoxy Resins.** a) Exotherms of a resin containing a 1:1.5 GDMP:BisDE molar ratio with 2 wt.% LC80 as a function of heating rate (1-10 °C min$^{-1}$). b) Flynn Wall Ozawa fits and calculated activation energies ($E_a$) for both the thiol-epoxy and epoxy-epoxy reactions.

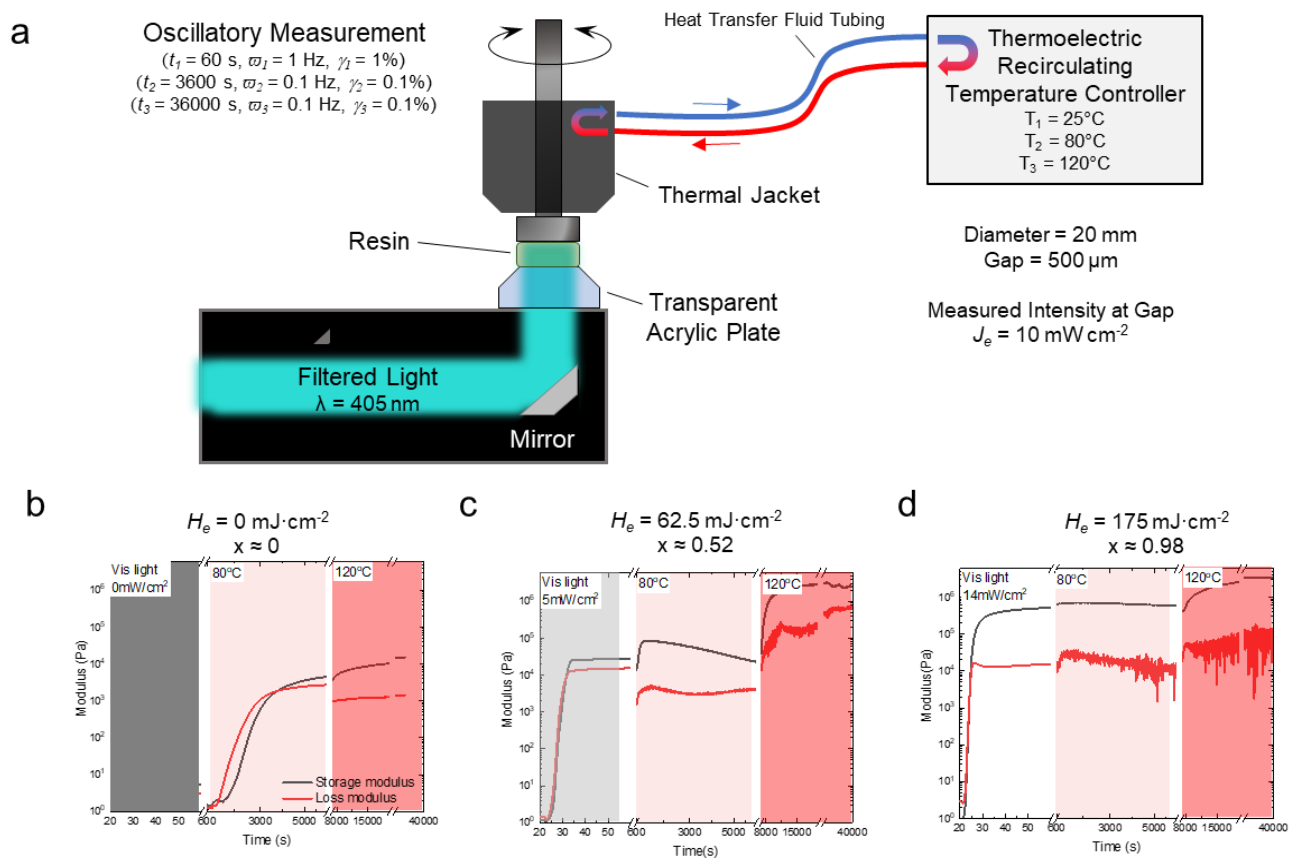

**Figure S5: Photo-thermal Rheology Experiments.** a) Rheological set-up to enable simultaneous photo- and thermal manipulation of the resin under curing. b-d) Change in storage ($G'$) and loss ($G''$) moduli over time under different initial photoexposure conditions.

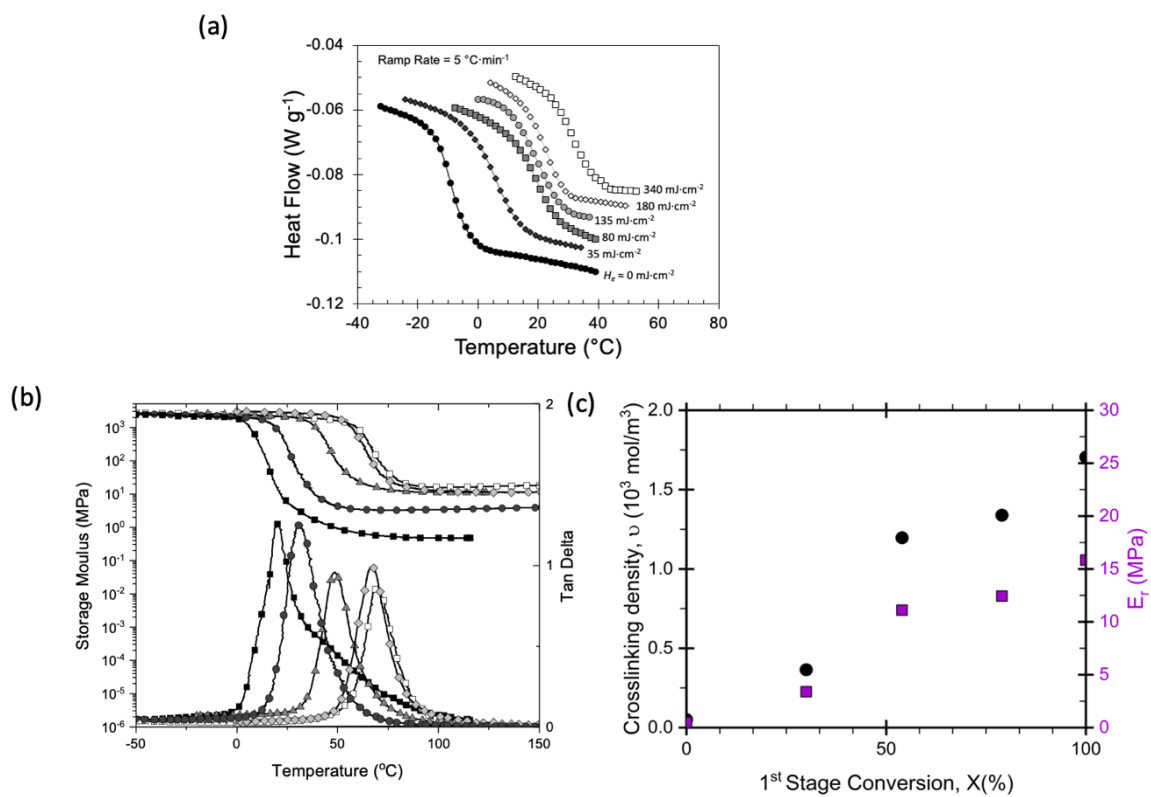

**Figure S6: (a) Differential Scanning Calorimetry (DSC) of Thiol-Ene-Epoxy Resins:** glass transition temperature exotherms for fully processed materials with different initial photodosages at a scanning rate of 5 °C min$^{-1}$. **(b) Dynamic Mechanical Analysis (DMA) of Thiol-Ene-Epoxy Resins:** storage modulus ($E$') and tan delta as a function of temperature for fully processed materials with different initial photodosage; (c) the crosslinking density and rubbery modulus versus 1$^{st}$ stage conversion. the crosslinking density is calculated based on $v=e_r/(3rt)$. Note, the glass transition temperatures obtained from DMA is higher than those from DSC, as previous demonstrated[4].

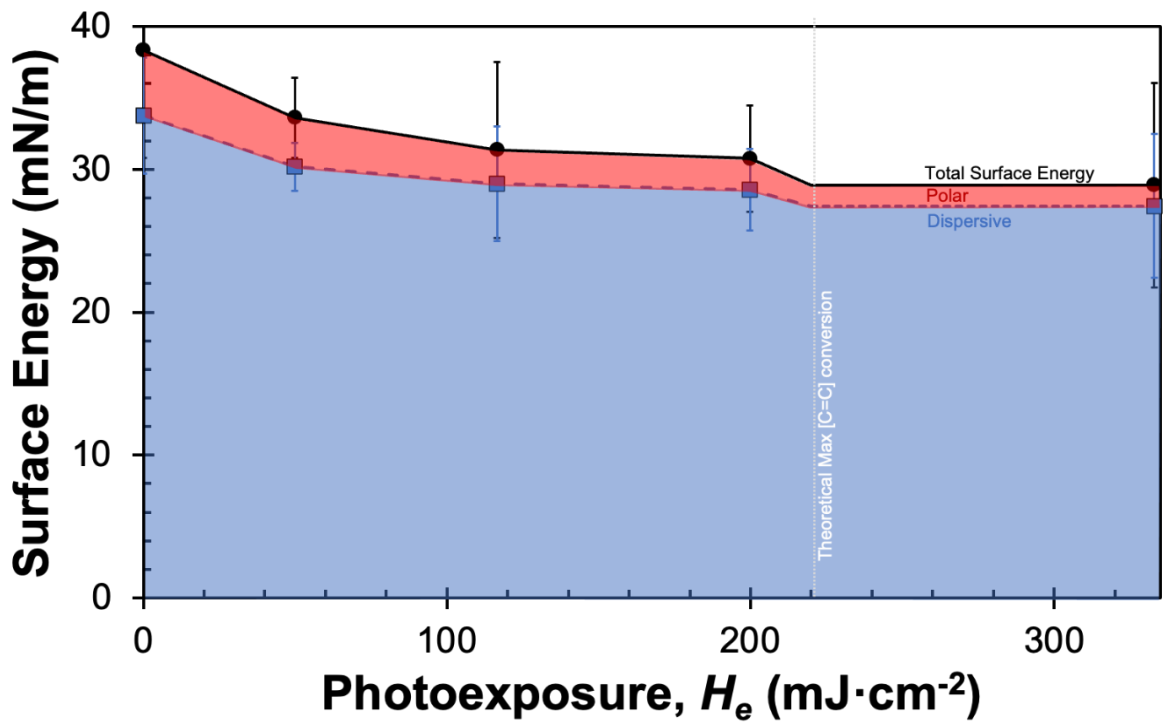

**Figure S7. Surface Energy for Fully Processed Materials with Different Initial Photodosages**

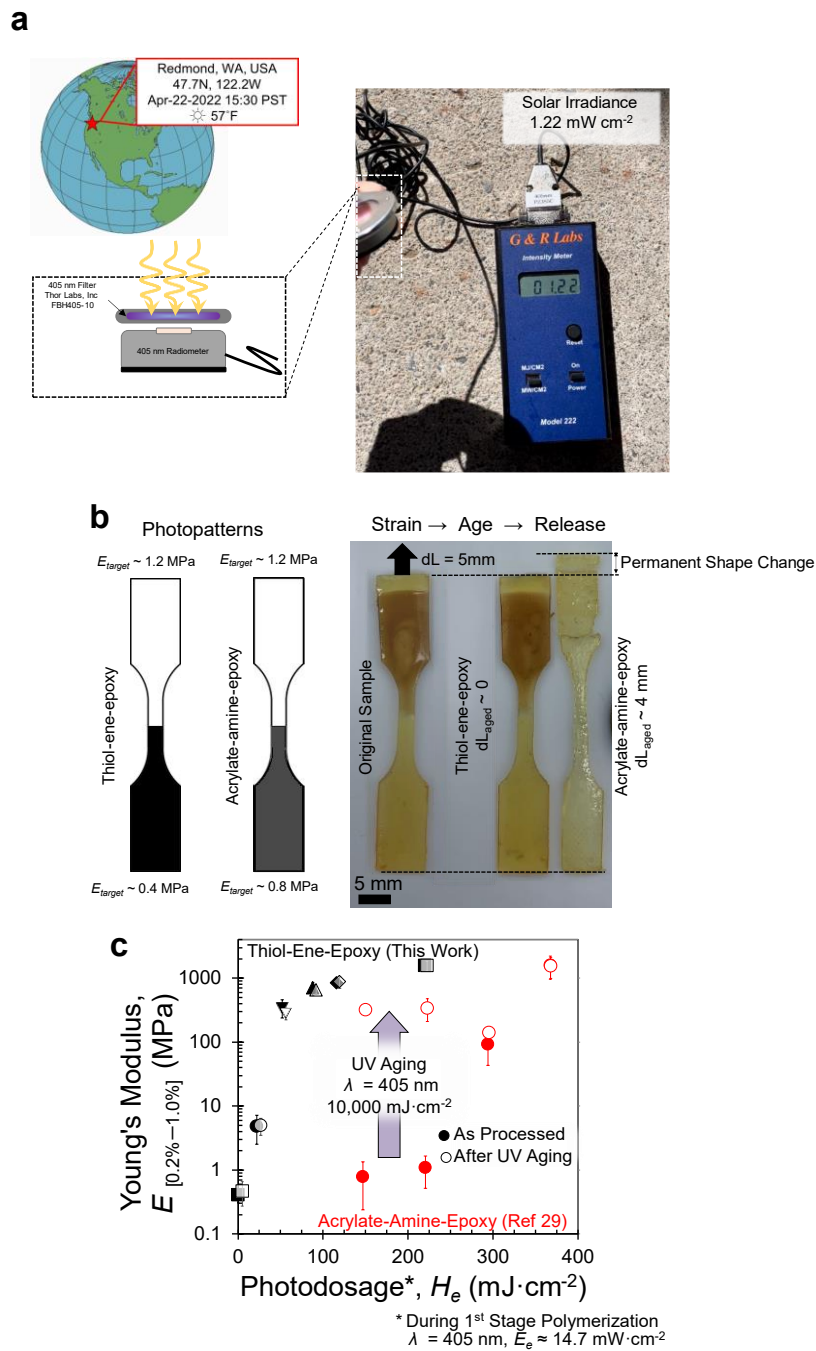

**Figure S8. Photostability of Ternary Photochemisrties.** a) Example solar irradiances for $\lambda$ = 405 nm can regularly exceed 1 mW cm$^{-2}$ even in Northern Latitudes (47.7 N). Data collected on April 22, 2022 at 15:30 PST. b) Target photopattern of Multimaterial (50/50) dogbone samples for both the thiol-ene-epoxy ($H_{e,initial}$ = 0 and 450 mJcm$^{-2}$) and acrylate-amine-epoxy ($H_{e,initial}$ = 150 and 450 mJcm$^{-2}$) systems. d) While subjected to a strain of dL = 5mm, samples were aged with an additional ~1500 mJ·cm$^{-2}$ of photoexposure. Prior to imaging samples were left for ~1h to allow for shape recovery. c) Average Young's Modulus as a function of initial photoexposure for the three stage multimaterial photochemistries (error bars represent standard deviation, N ≥ 7) both as processed and after UV aging ($H_{e,\,aged}$ = 10,000 mJ cm$^{-2}$, $\lambda$ = 405 nm). Unlike the acrylate-amine-epoxy system, our thiol-ene-epoxy chemistry exhibits near identical performance after aging (no statistically significant difference by unpaired t-test, see Table S2

for complete data). For clarity in reading data UV aged samples are offset by ($\Delta H_e = +5$ mJ cm$^{-2}$).

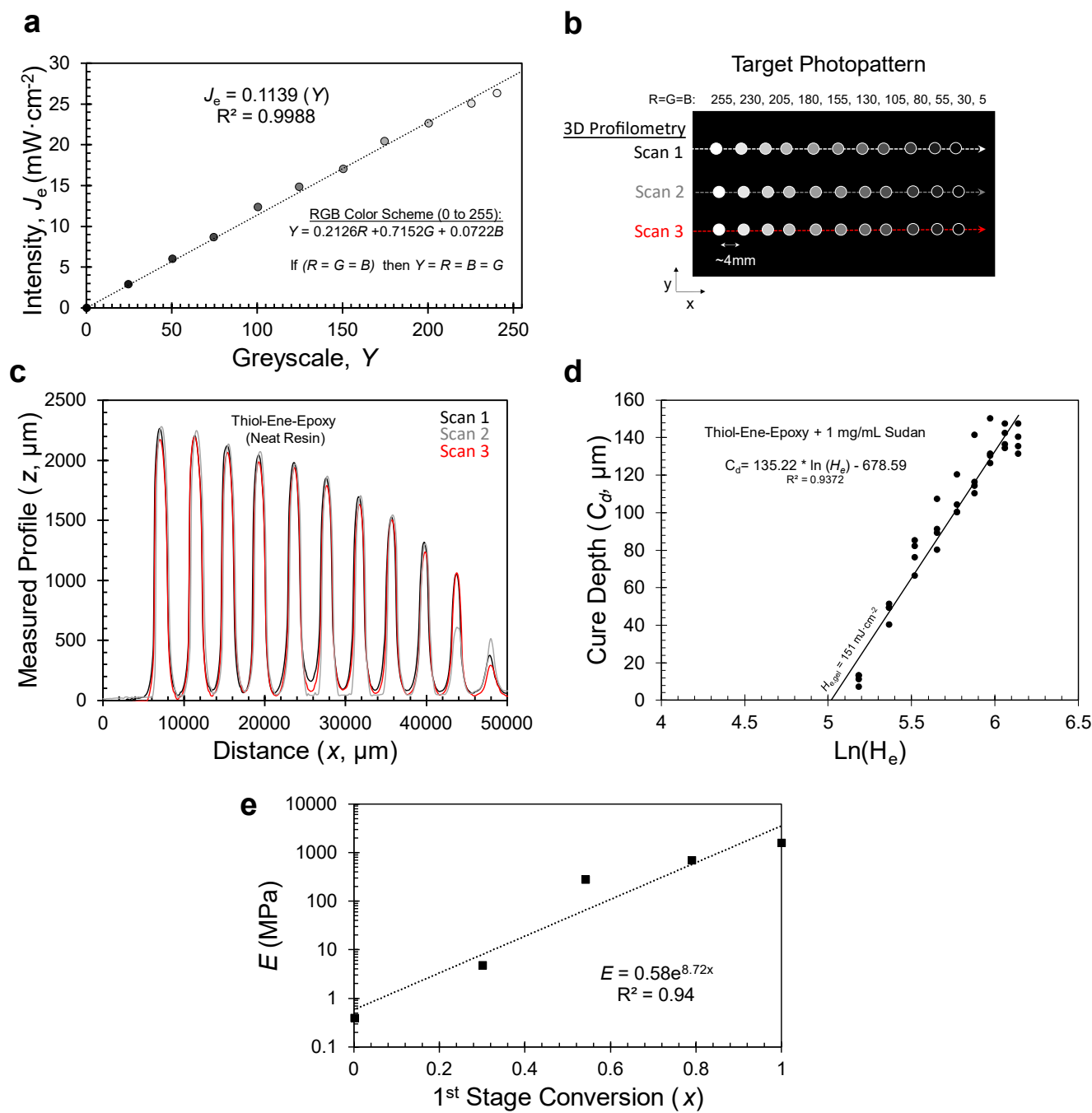

**Figure S9: Greyscale and Cure Depth Calibrations for Determining Print Parameters** a) Greyscale value vs. irradiative power for our printer. b) Target photopattern and 3D profilometry scans for cure depth experiments. c) Measured height profiles of the resulting cured pattern. d) Cure depth as a function of the natural logarithm of applied photodosage, Ln($H_e$) for our thiol-ene-epoxy resin with the addition of dye (1mg/1mL Sudan) which shows a different critical gel point than the pure resin in Main Text Figure 3b. e) Empirical fit of modulus as a function of 1$^{st}$ stage conversion.

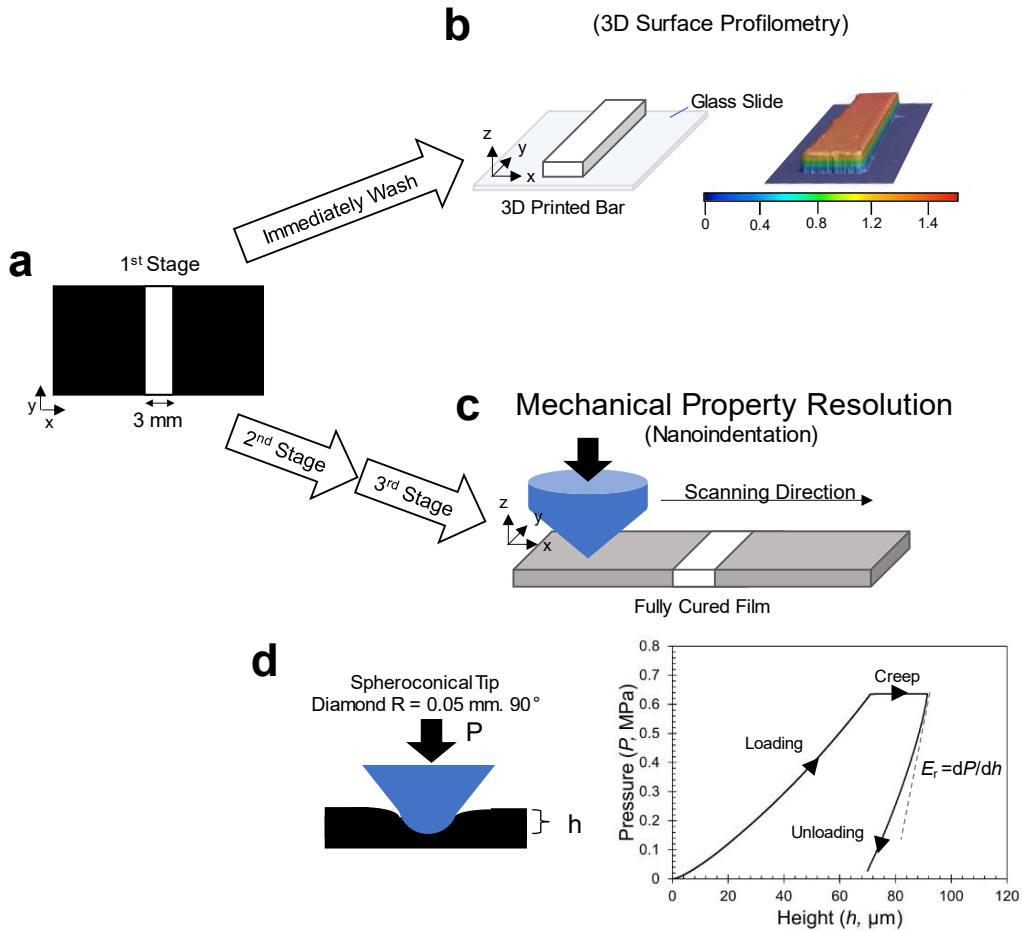

**Figure S10: Nanoindentation Experiment.** a) Applied photopattern to create samples for resolution characterization ($H_e = 0$ or 300 mJ·cm$^{-2}$). b) Immediate removal of unreacted precursors yields only the gelled material. 3D surface profilometry of the resulting structure allows for quantification of the shape. c) Fully processing the sample after photoexposure yields a solid film. d) With a spheroconical tip, we measure the pressure-displacement curves during nanoindentation and extract the reduced modulus, $E_r$, from the unloading curve.

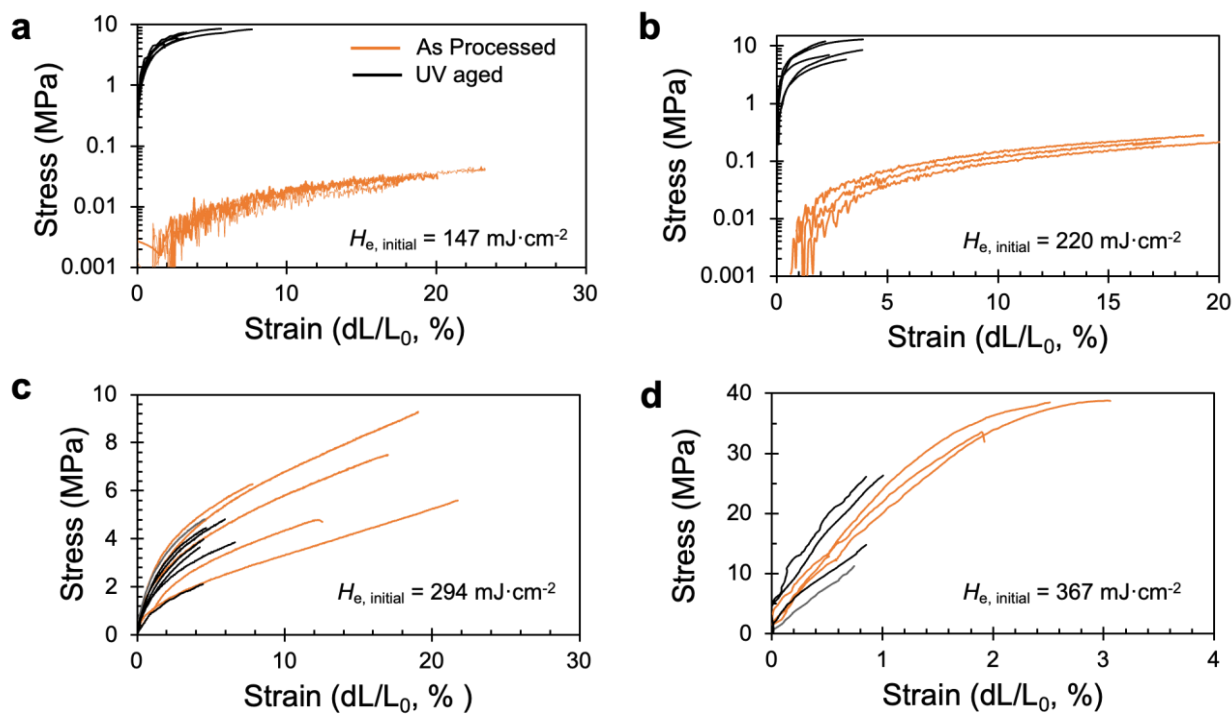

**Figure S11. Strain-Stress Curve of Acrylate-Amine-Epoxy System Before and After UV Aging.** The average stress-strain curves for acrylate-amine epoxy chemistry with initial photodosage of (a) 147 mJ·cm$^{-2}$; (b) 220 mJ·cm$^{-2}$; (c) 294 mJ·cm$^{-2}$; (d) 367 mJ·cm$^{-2}$